\documentclass[10pt]{article}
\usepackage{scicite}
\usepackage{times}
\usepackage[small,justification=justified]{caption} 
\usepackage{braket}
\usepackage[utf8]{inputenc}
\usepackage{hyperref}
\usepackage{amsmath} % Math
\usepackage{amssymb}  % Additional symbols
\usepackage{amstext}  % Text in equations
\usepackage{graphicx}
\usepackage{amssymb}
\usepackage{color}
\usepackage{subcaption}
\usepackage{mathrsfs}
\usepackage{physics}

\usepackage{setspace}
\usepackage{caption}
\let\oldbibliography\thebibliography
\renewcommand{\thebibliography}[1]{%
  \oldbibliography{#1}%
  \setlength{\itemsep}{-5pt}%
}

\newenvironment{sciabstract}{%
\begin{quote}}
{\end{quote}}

\title{Quantum interface of an electron and a nuclear ensemble} 

\author
{D. A. Gangloff,$^{1\ast\dagger}$ G.  \'Ethier-Majcher,$^{1\ast}$ C. Lang,$^{1}$ E. V. Denning,$^{1,2}$ J. H. Bodey,$^{1}$\\ D. M. Jackson,$^{1}$ E. Clarke,$^{3}$ M. Hugues,$^{4}$ C. Le Gall,$^{1}$ M. Atat\"ure$^{1\dagger}$\\
\\
\normalsize{$^{1}$Cavendish Laboratory, University of Cambridge,}\\
\normalsize{JJ Thomson Avenue, Cambridge CB3 0HE, United Kingdom}\\
\normalsize{$^{2}$Department of Photonics Engineering, Technical University of Denmark,}\\
\normalsize{2800 Kgs. Lyngby, Denmark}\\
\normalsize{$^{3}$EPSRC National Epitaxy Facility, University of Sheffield,}\\
\normalsize{Broad Lane, Sheffield, S3 7HQ, United Kingdom}\\
\normalsize{$^{4}$Universit\'e C\^ote d'Azur, CNRS, CRHEA,}\\
\normalsize{rue Bernard Gregory, 06560 Valbonne, France}\\
\\
\\
\normalsize{$^\ast$These authors contributed equally to this work}
\\
\normalsize{$^\dagger$To whom correspondence should be addressed:  dag50@cam.ac.uk, ma424@cam.ac.uk.}
}

\date{}
\begin{document}\sloppy 



\maketitle
\vspace*{20pt}
\doublespacing
\begin{sciabstract}
Coherent excitation of an ensemble of quantum objects underpins quantum many-body phenomena, and offers the opportunity to realize a quantum memory to store information from a qubit. Thus far, a deterministic and coherent interface between a single quantum system, e.g.~a qubit, and such an ensemble has remained elusive. We first use an electron to cool the mesoscopic nuclear-spin ensemble of a semiconductor quantum dot to the nuclear sideband-resolved regime. We then implement an all-optical approach to access these individual quantized electronic-nuclear spin transitions. Finally, we perform coherent optical rotations of a single collective nuclear spin excitation corresponding to a spin wave called a nuclear magnon. These results constitute the building blocks of a dedicated local memory per quantum-dot spin qubit and promise a solid-state platform for quantum-state engineering of isolated many-body systems.
\end{sciabstract}
%
%% 223 words
%
\twocolumn
\singlespacing

%\doublespacing
%\linespread{1.8}
%\vspace{5pt}
A controllable quantum system provides a versatile interface to observe and manipulate the quantum properties of an isolated many-body system\cite{Amico2008}. In turn, collective excitations of this ensemble can store quantum information as a long-lived memory\cite{Taylor2003a,Kurucz2009,Choi2010} -- one of the contemporary challenges in quantum technologies. This situation is captured elegantly by the central spin model\cite{Abragam1961,Stanek2014}, studied in donor atoms embedded in Si\cite{DeSousa2003,Pla2012}, in diamond color centers\cite{Childress2006,Balasubramanian2009,Kalb2017a}, and in semiconductor nanostructures\cite{Khaetskii2002,Merkulov2002,Bluhm2011,Urbaszek2013}. In these systems, the state of the central spin and of the spin ensemble that surrounds it are tied by mutual interaction, allowing proxy control over the many-body system in principle\cite{Tran2018}. Realising this scenario with an electron in a semiconductor quantum dot (QD) offers access to a dense ensemble of nuclear spins uniformly coupled to the central spin. In this system, coherent addressing of the ensemble via the central spin has yet to be shown, and a limiting factor is the thermal fluctuations of the surrounding spins that obfuscate the state-selective transitions required for such control. However, driving the central spin can stimulate exchange of energy with its surrounding spins, and thus modify the properties of its own environment. This has been shown to reduce the uncertainty on the collective spin state of the isolated QD nuclei, leading to prolonged electron spin coherence\cite{Stepanenko2006,Greilich2007a,Reilly2008,Xu2009,Vink2009,Bluhm2010,Issler2010,Chow2016,Onur2016,Ethier-Majcher2017}.

In this work, we use all-optical stimulated Raman transitions to manipulate the electron-nuclear system and realize a coherent interface. First employing a configuration analogous to Raman cooling of atoms\cite{Heinzen1990}, we drive the electron spin to reduce the thermal fluctuations of the nuclear spin ensemble (Fig.~1a).  Cooling the nuclear spin fluctuations to an effective temperature well below the nuclear Zeeman energy ($< 1$~mK) reveals an excitation spectrum of transitions between many-body states that are collectively enhanced by the creation of a single nuclear spin-wave excitation -- a nuclear magnon. Finally we drive a single magnon transition resonantly, inducing clear coherent exchange between the electron spin and the nuclear spin ensemble.
%310

% FIGURE 1
\begin{figure*}[!hbpt]
\begin{center}
   \includegraphics*[width=\textwidth]{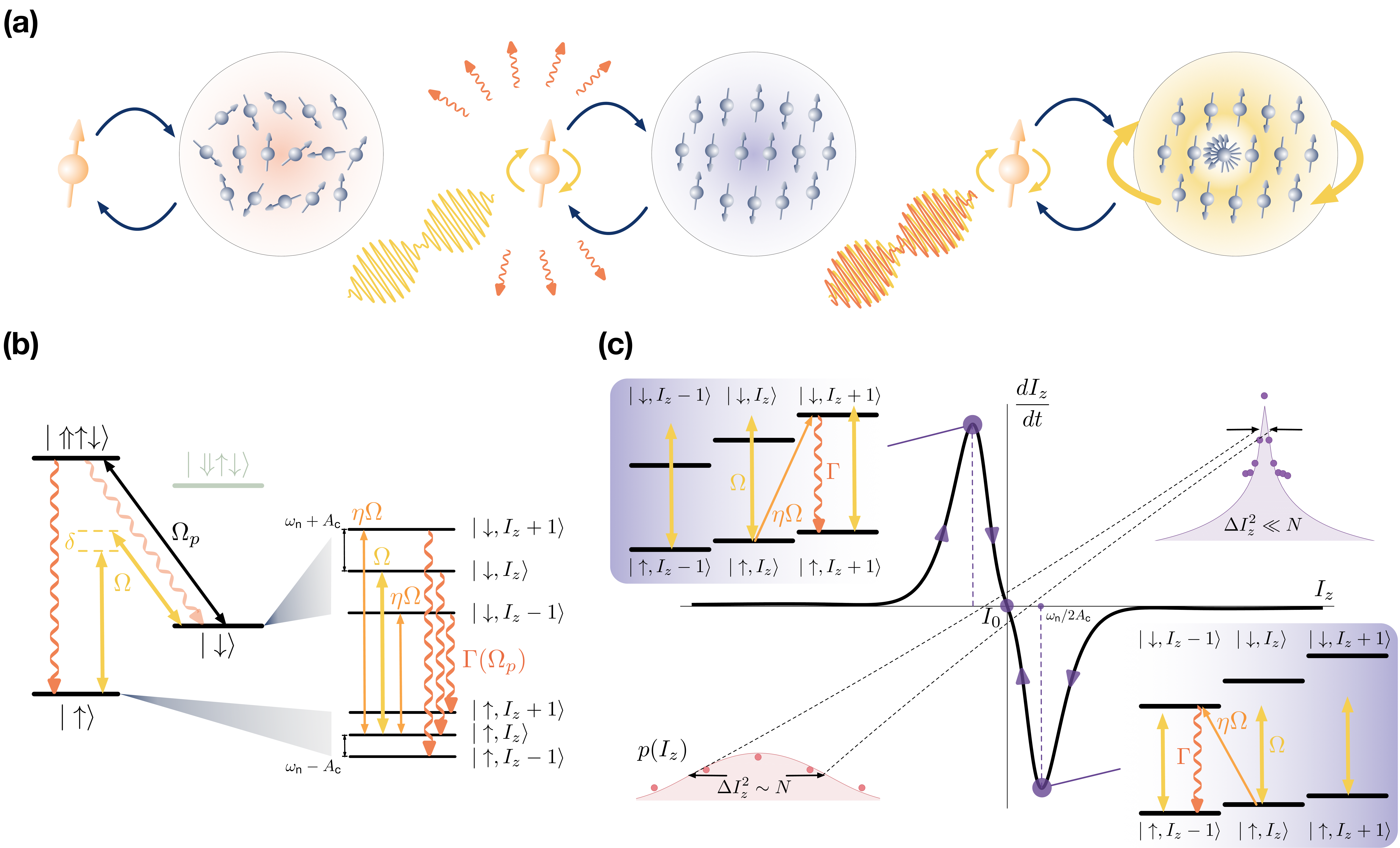}
             \centering
   \caption{\textbf{An electron controls a nuclear ensemble} \textbf{(a)} (left) Typical scenario for central spin problem where a spin interacts with a thermally fluctuating ensemble. (middle) When resonantly driven in the presence of dissipation, the spin can cool the ensemble to a lower effective temperature. (right) Driving the spin can create coherent superpositions of single spin-flips as collective excitations of the cooled ensemble. \textbf{(b)} Realization of the central spin problem through a semiconductor QD, under a magnetic field in Voigt geometry, optically pumped to electronic spin state $\ket{\uparrow}$ by a resonant drive $\Omega_p$ via the trion state $\ket{\Uparrow \uparrow \downarrow}$ of homogeneous linewidth $\Gamma_0 = 150$~MHz at a rate $\Gamma \sim \Omega_p^2/\Gamma_0 \leq 38$~MHz. The electron-spin splitting is (Overhauser) shifted by its hyperfine interaction $2A_\text{c} I_\text{z}$, where $A_\text{c} = 600$~kHz, with an ensemble of $N$ ($10^4$ to $10^5$) nuclear spins, described by mean polarization states $I_\text{z} = [-3N/2,3N/2]$ (taken for spin-$3/2$). Far-detuned ($\gtrsim 1$~nm) Raman beams drive the electron spin resonance (ESR) at a Rabi frequency $\Omega \lesssim 40$~MHz, including transitions that simultaneously flip a single nuclear spin $I_\text{z} \rightarrow I_\text{z}\pm1$ at frequency $\eta \Omega$ ($\eta < 1$). \textbf{(c)} Under a fixed frequency drive $\Omega$, the rate of change of nuclear polarization $dI_\text{z}/dt$ becomes a function of the polarization $I_\text{z}$, owing to the Overhauser shift and to the nuclear spin flipping transitions $W_\pm$ that reach a maximum rate when the drive is resonant with sideband transitions. The polarization $I_0$ is the stable point of this dynamical system, where the width $\Delta I_\text{z}^2$ of the nuclear polarization probability distribution $p(I_\text{z})$, measured by electron Ramsey interferometry\cite{SuppInfo}, is reduced (violet) compared to its value without cooling (red).}
       \label{fig:Fig1}
\end{center}
\end{figure*}
%245

% FIGURE 2
\begin{figure*}[!hbpt]
  	\begin{center}
   \includegraphics*[width=\textwidth]{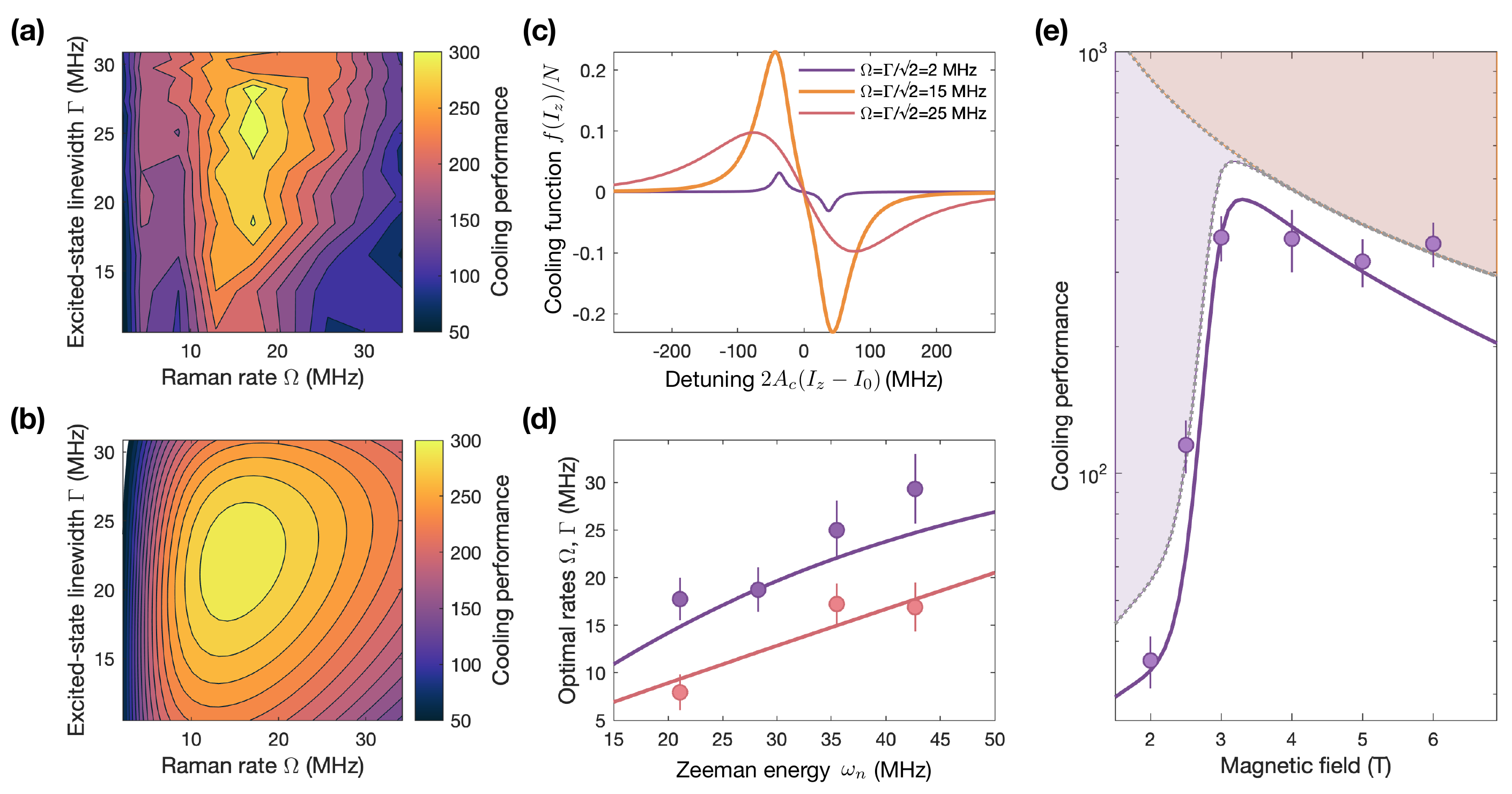}
             \centering
   \caption{\textbf{Optimal cooling of the nuclear ensemble} \textbf{(a)} Experimental Raman cooling performance $5N/4\Delta I_\text{z}^2$ as a function of Raman rate $\Omega$ and excited-state linewidth $\Gamma$, at $5$~T. The maximum of 300 is reached for $\Omega \sim \omega_\text{n}/2$, and saturation conditions $\Gamma \sim \sqrt{2}\Omega$. \textbf{(b)} Theoretical prediction of panel (a). \textbf{(c)} Calculated cooling curves $f(I_\text{z}) \propto W_+-W_-$ at optical saturation $\Omega=\Gamma/\sqrt{2}$ for increasing rates. The largest damping, $f'(I_0)$, and therefore the best cooling (orange curve) occurs when the rates are approximately half of the Zeeman splitting $\omega_\text{n} = 36$~MHz. \textbf{(d)} Raman rate and excited-state linewidth at the measured optimal cooling performance as a function of nuclear Zeeman splitting at $3$~T, $4$~T, $5$~T and $6$~T. Solid curves are the corresponding theoretical calculations. \textbf{(e)} Magnetic-field optimal cooling. Circles represent the maximum cooling performance at a given magnetic field. Curves are from a theoretical model\cite{SuppInfo}. The purple region is the cooling limit imposed by the electron dephasing from high frequency nuclear noise\cite{Bechtold2015,Stockill2016}. The red region is the limit intrinsic to Raman cooling based on the nuclear sidebands\cite{SuppInfo}. Including the effect of electron-mediated diffusion\cite{Wust2016} generates the solid curve. Error bars represent one standard deviation of uncertainty. }
       \label{fig:Fig2}
       \end{center}
\end{figure*}
%209

Our system consists of a charge-controlled semiconductor QD\cite{SuppInfo}, where a single electron spin is coupled to an isolated reservoir of $N$ ($10^4$ to $10^5$) nuclear spins magnetically, and to a charged exciton state optically (Fig.~1b). We drive the electron-nuclear system with a narrow two-photon resonance at detuning $\delta$ from an excited-state whose linewidth $\Gamma$ is tunable via the optical pumping rate of the electronic ground state (Fig.~1b), as with Raman cooling\cite{Heinzen1990}. The optical parameters set the dissipation rate relative to the energy scales relevant for cooling, which are the nuclear Zeeman energy $\omega_\text{n}$ and the hyperfine coupling energy per nucleus $A_\text{c}$, like the phonon and photon recoil energies for trapped atoms \cite{Monroe1995g}. In atomic physics, the motion of an atom relative to detuned driving fields leads to a velocity-dependent absorption rate via the Doppler effect and, together with the photon recoil momentum, to a damping force that is the basis of laser cooling of atomic motion\cite{Phillips1998}. In our system, the hyperfine interaction between the electron and nuclei leads to a shift of the electron spin resonance (ESR) that depends linearly on the net polarization $I_\text{z}$ of the nuclei\cite{Abragam1961}, the Overhauser shift $2A_\text{c} I_\text{z}$, and thus to a polarization-dependent absorption rate. In the presence of material strain, the hyperfine interaction enables optically driven nuclear spin flips that can be modeled as sidebands of amplitude $\eta \Omega$ ($\eta <1$) on a principal transition of amplitude $\Omega$ that flips the electron spin\cite{Huang2010,Hogele2012}. With fast electron spin reset, absorption on the sidebands at polarization-dependent rates $W_\pm(I_\text{z})$ can increase ($+$) or decrease ($-$) the mean nuclear polarization $I_\text{z}$, as shown in Fig.~1c, in a process known as dynamic nuclear polarization\cite{Abragam1961,Eble2006,Urbaszek2007,Tartakovskii2007,Maletinsky2007}. The evolution of this complex system pitting drift rates $W_\pm$ against a polarization-dependent diffusion rate $\Gamma_\text{d}(I_z)$ is captured elegantly by a simple rate equation\cite{Hogele2012,Yang2013}:
%283
\noindent
\begin{equation}
\begin{split}
\frac{dI_\text{z}}{dt} = -\frac{\Gamma_\text{tot}}{(3N/2)}\left[I_z - f(I_z)\right]\text{,}
\end{split}
\label{Eq:fb}
\end{equation}
\noindent
where $\Gamma_\text{tot} = W_+ + W_- + \Gamma_\text{d}$ is the total diffusion rate, and the cooling function $f(I_z) = (3N/2)(W_+ - W_-)/\Gamma_\text{tot}$ captures the bidirectional drift that reduces fluctuations, as in Doppler cooling\cite{Phillips1998}. Figure 1c displays the polarization-dependent rate of evolution (Eq.~1), offering $I_0 = \delta/(2A_\text{c})$ as the steady-state polarization of the dynamical system. The rate extrema occur when the Overhauser shift brings a sideband transition in resonance with the drive, $|2A_\text{c}(I_\text{z}-I_0)| \approx \omega_\text{n}$ (for $\omega_\text{n} \gg A_\text{c}$), suggesting that the polarization fluctuations can be contained well within the range defined by the nuclear Zeeman energy, $\omega_\text{n}$. The effective damping on the driven spin-$3/2$ ensemble is the cooling function gradient $(5/3)f'(I_0)$ at steady-state $I_0$. For a probability distribution $p(I_\text{z})$ arriving at steady-state under drift and diffusion forces, the fluctuations $\Delta I_\text{z}^2$ are reduced from their value $5N/4$ at thermal equilibrium (Fig. 1c) by\cite{Yang2013,SuppInfo}
\noindent
\begin{equation}
\begin{split}
\frac{\Delta I_\text{z}^2}{5N/4} = \frac{1-\left(\frac{2}{5N}I_0\right)^2}{1-\frac{5}{3}f'(I_0)}\text{.}
\end{split}
\end{equation}

We tie these reduced fluctuations $\Delta I_\text{z}^2$ to an effective ensemble temperature\cite{SuppInfo}. From the electron's perspective, a commensurate reduction of fluctuations occurs for a highly polarized nuclear ensemble, which to date has not been achieved. This occurs at thermal equilibrium when the energy $k_\text{B} T$ falls below the system's defining energy scale, here the nuclear Zeeman energy $\hbar \omega_\text{n}$. The fluctuations we observe in Fig.~1c thus correspond to an effective temperature below $T = \hbar \omega_\text{n} / k_\text{B} = 1$~mK.
%112

Figure 2 highlights the optimal conditions for cooling the nuclear ensemble. The decay of the electron's coherence over a time $T_2^*$ is a direct measure of the nuclear polarization fluctuations  $\Delta I_\text{z}^2 = 1/2(A_\text{c} T_2^*)^2$\cite{Ethier-Majcher2017,SuppInfo}. Ramsey interferometry on the electron spin\cite{Bechtold2015,Stockill2016} thus serves as our thermometer. We parametrize temperature as a cooling performance factor $(5N/4)/\Delta I_{z}^2$ as a function of Raman rate $\Omega$ and excited state linewidth $\Gamma$, as shown in Fig.~2a. Its maximum value of $\sim300$ is found where the Raman rate $\Omega = 17$~MHz is approximately half of the nuclear Zeeman splitting $\omega_\text{n}=36$~MHz, and the excited state linewidth corresponds to optical saturation, $\Gamma \sim 25$~MHz. This is in quantitative agreement with the theoretical prediction, shown in Fig.~2b, from a microscopic model that accounts for nuclear spin diffusion and inhomogeneous broadening\cite{SuppInfo}.
%151

The Raman rate $\Omega$ and the electronic excited-state linewidth $\Gamma$ are key parameters that determine the spectral selectivity and the diffusion rate of the cooling process. For best cooling, no absorption should occur at the stable point, while sideband absorption should turn on sharply in response to polarization fluctuations away from this stable point. This defines an optimal value for $\Omega$ and $\Gamma$ that relates to the sideband spacing $\omega_\text{n}$: $\Omega,\Gamma \ll \omega_\text{n}$ entails high spectral selectivity but weak sideband absorption near the stable point, while $\Omega,\Gamma \sim \omega_\text{n}$ entails strong  absorption on the sidebands but low spectral selectivity. As shown in Fig.~2c, this can be visualized in the dependence of the cooling function $f(I_\text{z})$ on the optical parameters. The damping $f'(I_0)$ is largest when the Raman rate is approximately half of the nuclear Zeeman energy, $\Omega \sim \omega_\text{n}/2$, and when it is close to saturation with the linewidth $\Omega \sim \Gamma/\sqrt{2}$. We confirm this experimentally in Fig.~2d by changing the applied magnetic field: the Raman rate and the excited-state linewidth that optimize the cooling performance are proportional to the sideband spacing. 
%174

The lowest temperature of our system is a function of distinct diffusion and broadening processes competing with Raman cooling, through magnetic-field dependent rates: in the low-field regime, homogeneous broadening of the ESR dominates\cite{Bechtold2015,Stockill2016} (blue region in Fig.~2e), while in the high-field regime optical diffusion does\cite{SuppInfo} (red region in Fig.~2e). Further, electron-mediated nuclear spin diffusion\cite{Latta2011,Wust2016} counteracts Raman cooling in both regimes. Figure 2e displays the magnetic field dependence of the temperature optimized against optical parameters. Our results follow closely the field-dependent bounds obtained from modelling the diffusion processes, and establish the globally optimal cooling performance of $\sim$$400$ at $\sim$$3.3$~T. Operating close to this field, we prepare the nuclear ensemble at an effective temperature of $200$~$\mu$K\cite{SuppInfo}. There, the Overhauser fluctuations are well below the nuclear Zeeman splitting, $2A_\text{c} \sqrt{\Delta I_\text{z}^2} =7$~MHz $< \omega_\text{n} = 22$~MHz (at 3~T), which places our system well into the sideband-resolved regime.
%170

%%% SIDEBANDS %%%%%

%FIGURE 3
\begin{figure*}[!hbpt]
  	\begin{center}
   \includegraphics*[width=\textwidth]{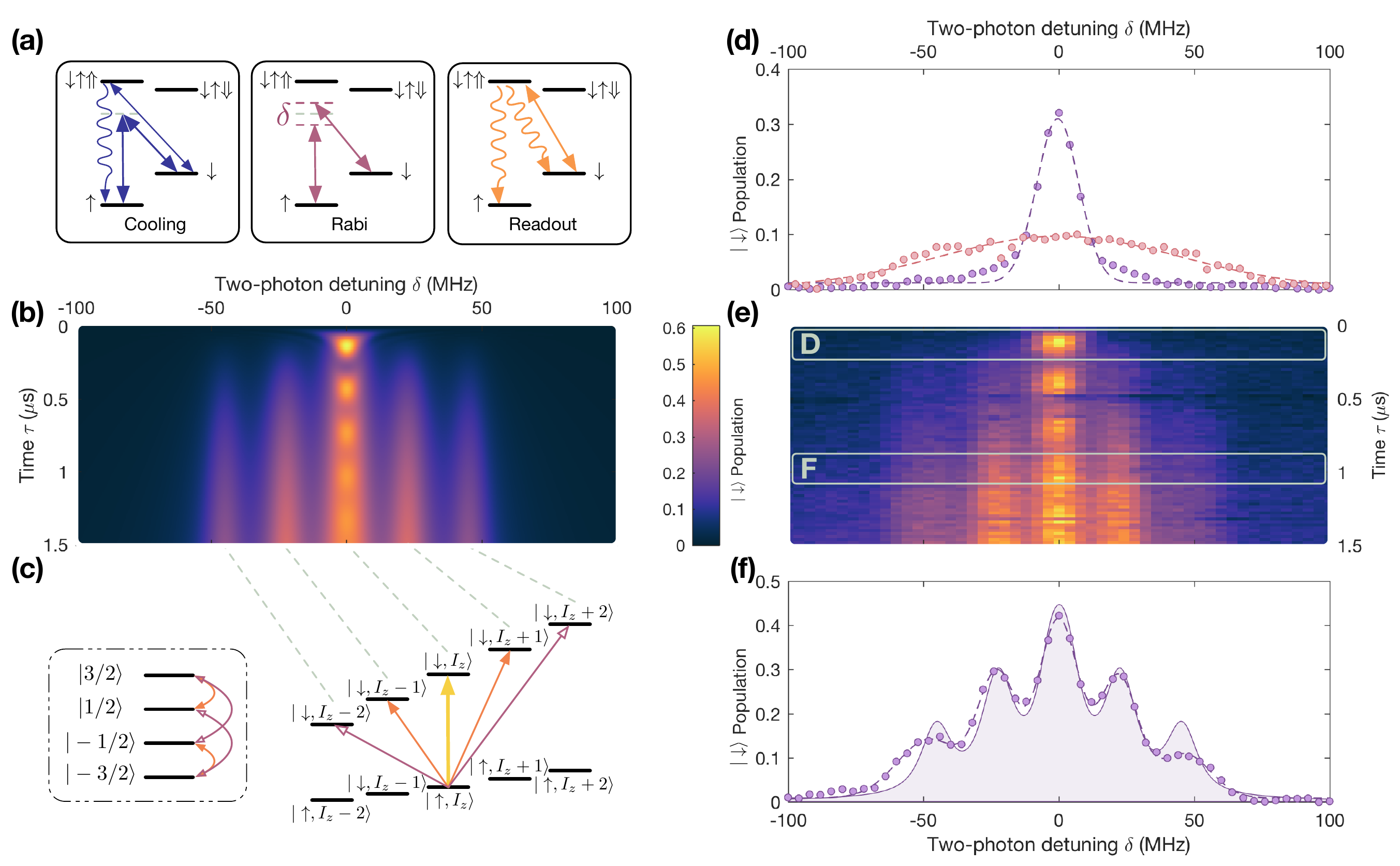}
             \centering
   \caption{\textbf{Resolving single nuclear magnons} \textbf{(a)} Spectrum measurement sequence, from left to right: Raman cooling, Rabi drive ESR at detuning $\delta$ for time $\tau$, and optical readout of the electron $\ket{\downarrow}$ population\cite{SuppInfo}. \textbf{(b)} Theoretical ESR spectrum buildup as a function of two-photon detuning $\delta$ and drive time $\tau$, at fixed nuclear polarization $I_\text{z} = 0$, for a Rabi frequency of $\Omega = 3.3$~MHz on the central transition. Sideband coupling $\eta$ is fitted\cite{SuppInfo}. \textbf{(c)} On the right, the ladder of electronic and nuclear states showing the carrier $I_\text{z} \rightarrow I_\text{z}$ and sideband transitions $I_\text{z} \rightarrow I_\text{z} \pm 1, I_\text{z} \pm 2$ from an initially spin-up polarized electron at a nuclear polarization of $I_\text{z}$. On the left, the same transitions represented within a single nuclear spin-$3/2$ manifold. \textbf{(d)} Spectra with optimal (violet) and poor (red) Raman cooling at average delay $\tau = 0-150$~ns. The dashed curves are Gaussian fits with standard deviation $7.7$~MHz and $44.6$~MHz, respectively. \textbf{(e)} Experimental spectrum buildup with $\Omega = 3.8$~MHz. \textbf{(f)} Spectrum at integrated delay $\tau = 850-1000$~ns. The solid curve is the same time slice averaged from the theory spectrum of panel (b). The dashed curve is five Gaussian functions centered at $\delta \sim 0,\pm\omega_\text{n},\pm 2\omega_\text{n}$\cite{SuppInfo}.}
       \label{fig:Fig3}
       \end{center}
\end{figure*}
%154

%164
We now probe the electron-spin state in the coherent regime where dissipation is turned off, $\Gamma \rightarrow 0$. We drive the ESR for a time $\tau$ at a detuning $\delta$ and measure the electron $\ket{\downarrow}$ population (Fig.~3a). Figure 3b shows a spectrum buildup over time $\tau$, obtained from our theoretical analysis\cite{SuppInfo}, where we expect five distinct processes, as shown in Fig. 3c: a central transition at $\delta=0$, and four sideband transitions at $\delta = \pm\omega_\text{n},\pm 2\omega_\text{n}$. The model consists of a master equation treatment of the driven electron-nuclear system that accounts for electron dephasing, where the nuclear system is reduced to collective states with polarization close to $I_0$\cite{SuppInfo}. The microscopic origin of the nuclear spin-flip sideband transitions is the strain-induced electric field gradient that couples to the quadrupole moment of the quantum-dot nuclei with spin $I \geq 3/2$, which mixes their Zeeman eigenstates\cite{Urbaszek2013}. From a first-order perturbative expansion of the hyperfine interaction\cite{SuppInfo}, the ESR is dressed by sideband transitions that change the nuclear polarization by one quantum ($I_\text{z} \rightarrow I_\text{z}\pm1$)\cite{Huang2010,Hogele2012} and two quanta ($I_\text{z} \rightarrow I_\text{z}\pm2$), with comparable strength. When the driving field with amplitude $\Omega$ is detuned from the principal transition by one or two units of nuclear Zeeman energy $\omega_\text{n}$, these resonant transitions occur with an amplitude $\eta \Omega$, as sidebands of strength $\eta = \mathcal{D} A_{\text{nc}}/\omega_{\text{n}}$; here $A_{nc} \approx 0.015 A_\text{c}$ is the non-collinear hyperfine constant parametrizing the perturbation. The driven electron cannot distinguish the $\sim N$ possible spin-flips that take $I_z$ to $I_z \pm 1, \pm 2$, which leads to the degeneracy factor $\mathcal{D}  \sim \sqrt{N}$. This underpins the collective enhancement\cite{Dicke1954b} that makes the nuclear spin-flip sideband transitions so prominent in our system.

Figure~3d shows the experimental spectra averaged over short delays $\tau=0-150$ns, where $\Omega \tau \sim \pi$, revealing the principal ESR with optimal (violet data) and suboptimal (red data) cooling. The feature width is a convolution of the drive Rabi frequency $\Omega$ with the Overhauser field fluctuations $2A_\text{c} \sqrt{\Delta I_\text{z}^2}$, and highlights the spectral narrowing achieved by Raman cooling. Figure 3e shows the time-frequency map of this measurement. At $\delta=0$, the principal ESR leads to Rabi oscillations at $\Omega = 3.8$~MHz. At larger delays where $\eta \Omega \tau \sim \pi$ and at a sufficient detuning from the principal transition $\delta \gg \Omega$, the emergence of four sideband processes agrees well with our predictions. Figure 3f is a standout observation of the sideband spectrum, integrated over $\tau=850-1000$ns. A five-Gaussian fit (dashed curve) verifies that the sidebands emerge at integer multiples of $\omega_n$, and the shaded area highlights the theoretical spectrum. Our results confirm that the sideband drive can excite selectively a single nuclear spin-flip in the ensemble and highlight that $\sim N$ sufficiently identical nuclei are simultaneously coupled to the driven electron. Until now, such a collective nuclear-spin excitation had only been observed as ensemble measurements of atomic gases\cite{Johnson1984} and magnetic materials\cite{Seewald1997,Abdurakhimov2015}, while our result represents the deterministic generation of a single nuclear magnon by interfacing the nuclei with an elementary controllable quantum system.
%289

%FIGURE 4
\begin{figure}[!h]
  \centering
  \includegraphics[width=250pt]{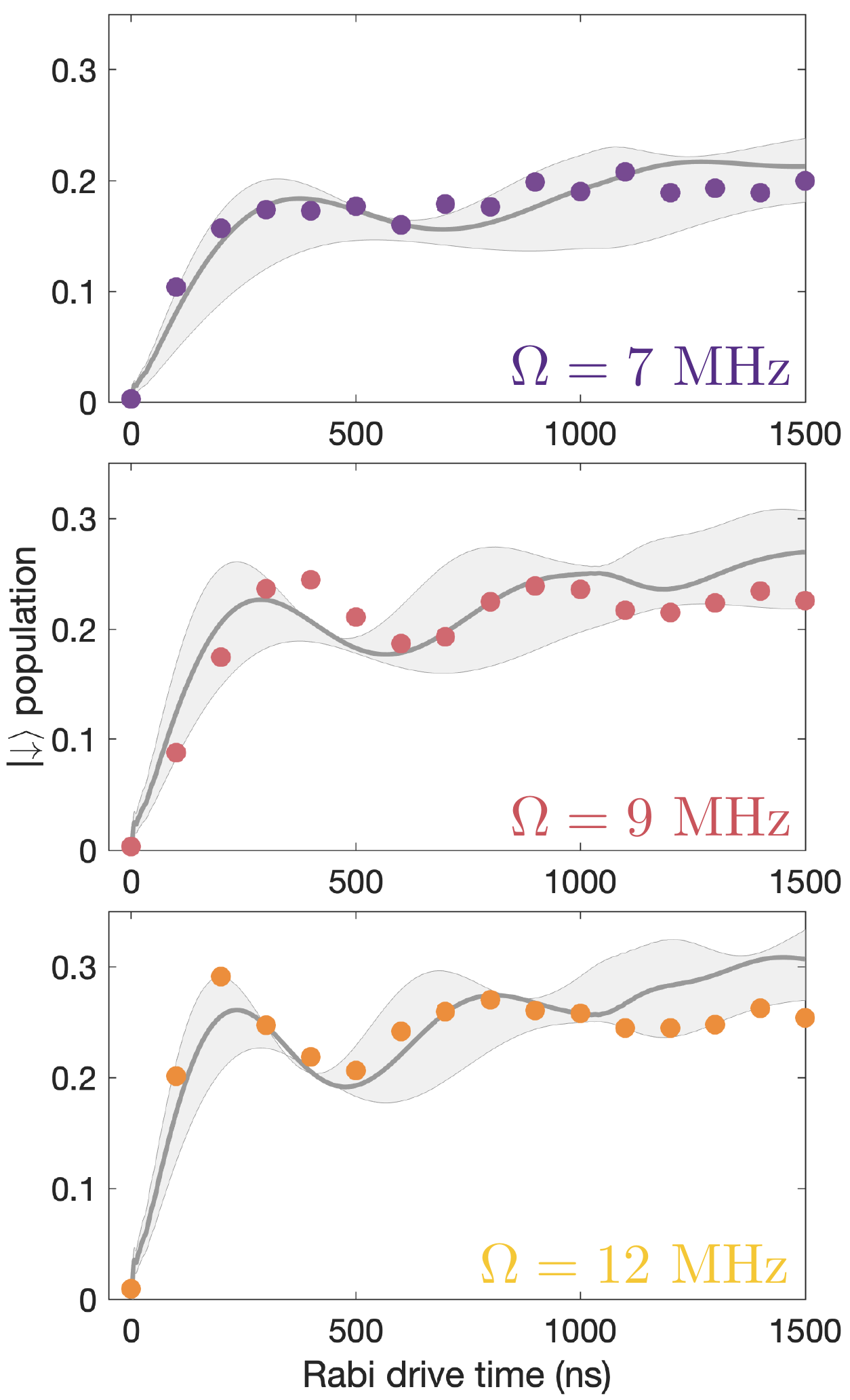}
  \caption{\textbf{Coherent oscillations of a nuclear magnon} Electronic excited-state $\ket{\downarrow}$ population\cite{SuppInfo}, measured after a Rabi pulse of $\tau$ at $\delta =-2\omega_\text{n} = -52$~MHz detuning, at 3.5T. The carrier Rabi frequency $\Omega$ is $7,9,$ and $12$~MHz for measurements shown in the top, middle, and bottom panels, respectively. Solid curves are the corresponding theoretical calculations with $\eta = 15\%$, using the same carrier Rabi frequencies. The shaded areas represent a $\sim\pm20\%$ deviation in model Rabi frequency.}
\end{figure}
%57

This spectral selectivity enables coherent generation of a single-spin excitation, provided it is faster than the dephasing times of the electron ($T_2 \approx 1$~$\mu$s\cite{Stockill2016}) and the nuclei ($\approx 10$~$\mu$s\cite{Wust2016}). Figure 4 illustrates this coherent drive via Rabi oscillations. We drive the $I_\text{z} \rightarrow I_\text{z} + 2$ sideband with $\eta \Omega > 1/T_2$ (Fig.~3c), and measure for delays $\tau \gtrsim \pi/\eta\Omega$. Figure 4 presents measurements with three Rabi frequencies $\Omega=7,9,12$~MHz. Oscillations of the electron spin population at a fraction $\eta$ of the carrier frequency $\Omega$ are a direct measurement of coherent electron-nuclear dynamics. We attribute the sharp appearance of oscillations as a function of Rabi frequency to reaching a sufficient sideband coupling $\eta \Omega$ to overcome inhomogeneities. The master equation model (solid lines in Fig.~4) captures this inhomogeneous broadening that limits the Rabi oscillations. The grey-shaded areas represent $\pm 20\%$ deviations of Rabi frequency, and our data's drift towards lower Rabi frequency at long delays suggests a dephasing mechanism that depends on accumulated phase $\Omega \tau$. Our model further allows us to reconstruct the nuclear-spin population transfer, where the effect of off-resonant excitation of the principal transition is not present, and shows that the electron spin population transfer is accompanied predominantly by nuclear spin population transfer\cite{SuppInfo}.  
%216

The value $\eta \sim 15\%$, directly extracted from the coherent oscillations in Fig.~4, confirms the $\sim \sqrt{N}$ enhancement of the sideband transition strength arising from the collective nature of the magnon excitation. Indeed, owing to sufficient coupling homogeneity, the nuclei can be treated as an ensemble of $N=30,$$000$ indistinguishable spins under the hyperfine interaction with the electron. Oscillations in Fig.~4 indicate the creation and retrieval of a coherent superposition of a single nuclear spin excitation among all spins, forming the basis of many-body entanglement as found for Dicke states\cite{Dicke1954b}. This occurs despite operating near zero polarization, where the degeneracy of nuclear states is maximal. Strikingly, this exchange of coherence is far from the bosonic approximation available for a fully polarized ensemble\cite{Taylor2003a}. Furthermore, an intermediate drive time $\eta\Omega\tau = \pi/2$ generates an inseparable coherent superposition state for the electron and the nuclei.

%%%% CONCLUSIONS

In this work, we have realized a coherent quantum interface between a single electron and $30$,$000$ nuclei using light. Making use of the back-action of a single nuclear-spin flip on the electron, the development of a dedicated quantum memory per electron spin qubit in semiconductor QDs becomes viable. Future possibilities also include creating and monitoring tailored collective quantum states of the nuclear ensemble, such as Schr\"{o}dinger cat states, by harnessing Hamiltonian engineering techniques. 
%122

\section*{Acknowledgments}
We thank Andreas Nunnenkamp and Guido Burkard for critical reading of the manuscript. We thank Robert Stockill for helpful discussions. This work was supported by the ERC PHOENICS grant (617985) and the EPSRC Quantum Technology Hub NQIT (EP/M013243/1). D.A.G. acknowledges support from a St John's College Title A Fellowship, G.\'E.-M. from the NSERC Postdoctoral Fellowship program, and E.V.D. from the Danish Council for Independent Research (Grant No. DFF-4181-00416).

\clearpage
\renewcommand{\thefigure}{S\arabic{figure}}
\setcounter{figure}{0}
\part*{Supplementary Information}

\tableofcontents

\section{Experimental System}
\subsection{QD Device}
Self-assembled InGaAs QDs are grown by Molecular Beam Epitaxy (MBE) and integrated inside a Schottky diode structure\cite{Urbaszek2013b}, above a distributed Bragg reflector to maximize photon outcoupling efficiency. There is a 35-nm tunnel barrier between the n-doped layer and the QDs, and a tunnel barrier above the QD layer to prevent charge leakage. The Schottky diode structure is electrically contacted through Ohmic AuGeNi contacts to the n-doped layer and a semitransparent Ti gate (6 nm) is evaporated onto the surface of the sample. The photon collection is enhanced by placement of a superhemispherical cubic zirconia solid immersion lens (SIL) on the top Schottky contact of the sample. We estimate a photon outcoupling efficiency of 10\% for QDs with an emission wavelength around 950 nm.

\subsection{Raman laser system}

Raman beams are generated by modulating a fiber-based EOSPACE electro-optic amplitude modulator (EOM) with a Rohde \& Schwartz 22GHz microwave frequency source amplified to a peak power of $25$~dBm. The Raman beams, with a combined optical power of $\sim$$1$~mW at the entrance to the cryostat, are passed through a quarter-wave plate and arrive at the quantum dot with near circular polarization at a red single-photon detuning $\Delta \gtrsim 2$~nm. The first-order EOM sidebands are two coherent laser fields whose energy difference can be made resonant with the electron spin resonance, as per this work $13-38$~GHz (corresponding to an applied magnetic field $B = 2-6$~T), leading to a two-photon detuning $\delta \approx 0$.

\section{Overhauser fluctuation measurements via electron spin coherence, $T_2^*$}

% FIGURE S1
\begin{figure}[!b]
  \centering
  \includegraphics[width=\columnwidth]{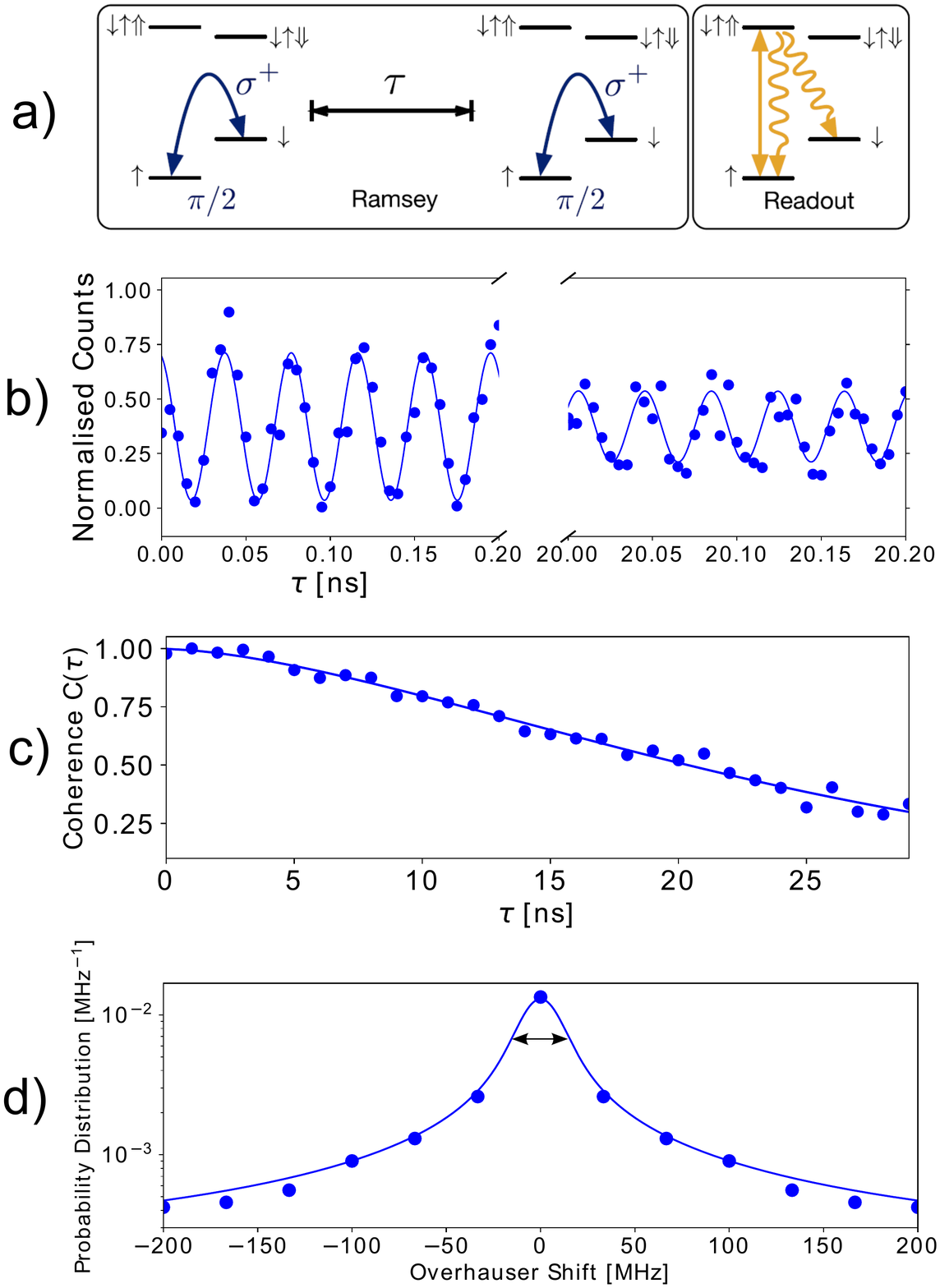}
  \caption{\textbf{Measuring nuclear spin fluctuations.} Data is taken at a magnetic field of $4$~T. \textbf{(a)} Ramsey interferometry experimental sequence. \textbf{(b)}  Normalized readout fluorescence as a function of Ramsey delay $\tau$. The solid curves are sinusoidal fits. \textbf{(c)} Coherence function $C(\tau)$ defined as the visibility of the sinusoidal fit to the readout data in (b). The solid curve is a stretched exponential function $\exp(-(\tau/T_2^*)^\alpha)$, with fitted parameters $T_2^* = 26\pm7$~ns and $\alpha = 1.6\pm0.2$. \textbf{(d)} Probability distribution of the Overhauser shift, obtained from the Fourier transform of the coherence function $C(\tau)$.}
\end{figure}

The electron coherence function $C(\tau)$ is measured by performing Ramsey interferometry on the electron spin\cite{Bechtold2015b,Stockill2016b}. To do so, we make use of a mode-locked Coherent MIRA ps-pulsed laser detuned from the trion state resonance by $\sim$$2$~nm to perform ultrafast $\pi/2$ rotations of the electron spin\cite{Press2008b,Greilich2009b}, separated by a time delay $\tau$, followed by a spin-selective resonant readout, as shown in Fig.~S1a. The visibility of the Ramsey fringes as a function of delay time $\tau$ (Fig.~S1b) is captured by the coherence function $C(\tau)$ (Fig.~S1c). The envelope of $C(\tau)$ is related by a simple Fourier transform to the Overhauser shift probability distribution $p(2A_\text{c} I_\text{z})$ (Fig.~S1d)\cite{Onur2018b}, whose variance $4A_\text{c}^2 \langle\Delta I_\text{z}^2 \rangle = 2/T_2^{*2}$ is our measure of the effective nuclear ensemble temperature:
\begin{equation*}
\begin{split}
C(\tau) = \left|\int_{-\infty}^{\infty} p(2A_\text{c}I_\text{z}) \exp\left(-i2A_\text{c} I_\text{z} \tau \right)d(2A_\text{c}I_\text{z})\right|
\end{split}
\label{Eq:fb}
\end{equation*}

\section{Electron Hahn Echo $T_2$}

The homogeneous dephasing time of the electron $T_2$ sets an important limit on cooling at low fields (Fig. 2e main text), and is one limit for coherent electron-nuclear exchange at the optimal field (Fig.~4 main text). Dephasing is dominated by high-frequency noise near the nuclear Zeeman frequencies arising from the hyperfine coupling of the electron to the nuclei, whose Zeeman eigenstates are mixed by quadrupolar interactions\cite{Bechtold2015b,Stockill2016b}. In Fig.~S2, we present measurements of the Hahn Echo $T_2$\cite{Press2010b} measured at magnetic fields of $2$~T, $3$~T, and $5$~T. These measured values are those used in our Raman cooling theory of Fig.~2 (main text) as well as the theory spectrum of Fig.~3 (main text).

% FIGURE S2
\begin{figure}[!t]
  \centering
  \includegraphics[width=\columnwidth]{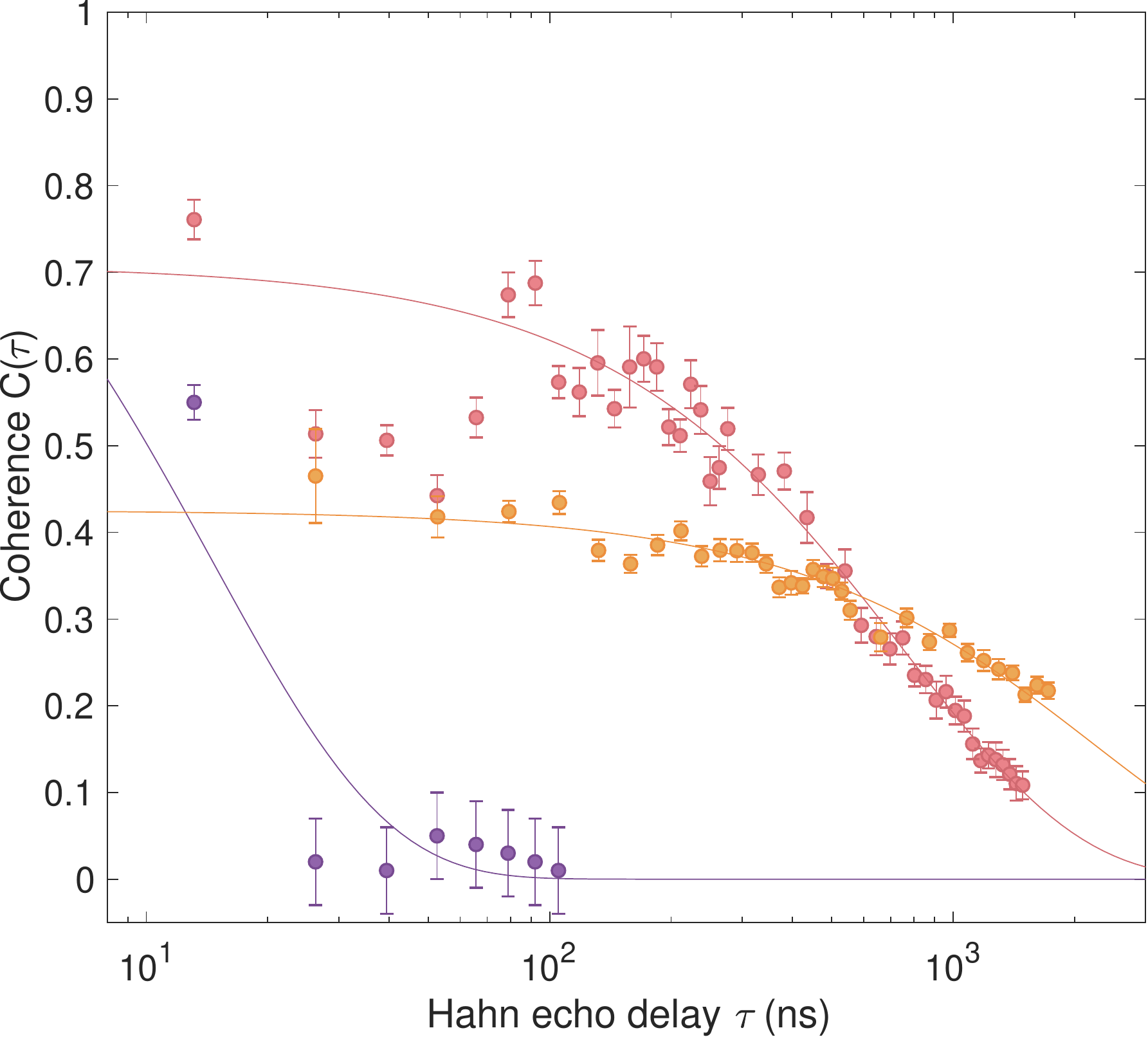}
  \caption{\textbf{Hahn Echo $T_2$ measurements} Measurement of the electron coherence as a function of the Hahn echo delay $\tau$ (time between two Ramsey $\pi/2$ pulses, with a $\pi$ pulse inserted at $\tau/2$) at magnetic fields of $2$~T (purple), $3$~T (red), and $5$~T (orange). The data are fitted with a simple exponential model $A\exp(-t/T_2)$, with the following fitted $T_2$ values: $15 \pm 10$~ns ($2$~T), $765 \pm 15$~ns ($3$~T), $2220 \pm 70$~ns ($5$~T).}
\end{figure}

\section{Note on electron excited-state population}

In Figs.~3 and 4 (main text) we present measurements of the electron excited-state population, which we perform experimentally by measuring the average trion fluorescence following a spin-selective resonant laser pulse. For best agreement with our theoretical analysis in Figs. 3 and 4 (main text), we have found that we needed to leave the conversion factor from readout fluorescence to population as a free parameter. In both figures, this conversion factor is fitted to $60\%$ of the value that would be expected from taking the resonantly driven electron population to be $0.5$ at driving times well beyond its coherence time $T_2$.

\section{A note on multiple nuclear species}

We expect that all three species, As, In, and Ga, are engaged in nuclear spin-flip processes that change the ensemble polarization $I_z$, and that the fluctuations $\Delta I_z^2$ arising from all three species are reduced by cooling, albeit perhaps unequally. Nonetheless, for overall simplicity and to reduce the number of free parameters in our models, we make a single-species approximation under which As nuclei dominate the dynamics. This is justified in several ways: 

\begin{itemize}
\item the quadrupolar angle of As makes it the species with the largest $A_\text{nc}$ value\cite{Bulutay2012b}
\item As has the lowest Zeeman energy, which makes its sideband coupling value of $\eta \propto \omega_\text{n}^{-2}$ the largest
\item In the case of cooling, absorption on the first sideband $I_z \pm 1$ of the species with the lowest Zeeman energy is the first transition to become resonant for a deviation away from the steady-state.
\end{itemize}

\section{Microscopic origins of the sideband transitions}

This section describes the perturbed hyperfine interaction that is the microscopic origin of the nuclear spin-flip sideband transitions.

The full Hamiltonian of the Raman-driven system with the Voigt $B$-field taken along the $z$-axis is (in a frame rotating with the Raman lasers and with the trion adiabatically eliminated)
\begin{align*}
  H=\delta S_\text{z} + 2\Omega S_\text{x} + \sum_j\omega_\text{n}I_\text{z}^j - \sum_j 2A_\mathrm{c}I_\text{z}^jS_\text{z} + H_\text{Q},
\end{align*}
where we have ignored the flip-flop terms in the hyperfine interaction and $H_\text{Q}$ describes the nuclear quadrupolar interaction \cite{Hogele2012b},
\begin{align*}
\begin{split}
  H_\text{Q}=\sum_j &B_\text{Q}\Big[I_\text{x}^{j2}\cos^2\theta + I_\text{z}^{j2}\sin^2\theta   +(I_\text{x}^jI_\text{z}^j+I_\text{z}^jI_\text{x}^j)\frac{\sin2\theta}{2}\Big]\text{,}
\end{split}
\end{align*}
\noindent
where $\theta$ is the angle between the z-axis and the quadrupolar axis.
Since the nuclear Zeeman splitting is much larger than the quadrupolar interaction strength, $B_\text{Q}$, the terms in $H_\text{Q}$ that do not commute with $I_\text{z}$ describe processes that are not allowed to first order. First, we extract the $I_\text{z}$-preserving and non-preserving parts of $H_\text{Q}$, which we denote by $H_\text{Q}^0$ and $V_\text{Q}$,
\begin{align*}
\begin{split}
  H_\text{Q}^0&=\sum_j B_\text{Q}[I_\text{z}^{j2}\sin^2\theta+\frac{1}{2}(I_\text{x}^{j2}+I_\text{y}^{j2})\cos^2\theta]\\
V_\text{Q}&=\frac{1}{2}\sum_j B_\text{Q}[(I_\text{x}^{j2}-I_\text{y}^{j2})\cos^2\theta
+(I_\text{x}^jI_\text{z}^j+I_\text{z}^jI_\text{x}^j)\sin2\theta].
\end{split}
\end{align*}
The $I_\text{z}$-commuting part, $H_\text{Q}^0$, induces an anharmonic shift of the single-nucleus spin ladder of $\Delta_\text{Q}=B_\text{Q}(2\sin^2\theta-\cos^2\theta)$, whereas $V_\text{Q}$ couples the nuclear Zeeman states.
To obtain an effective Hamiltonian for the allowed low-energy excitations, we dress $H$ with $V_\text{Q}$ using a Schrieffer-Wolff transformation \cite{Schrieffer1966b,Bravyi2011b}. This removes $V_\text{Q}$ and instead leads to a second-order correction term in the hyperfine interaction,
\begin{align*}
\begin{split}
  V_\text{Q}'=S_\text{z}\sum_j&A_\mathrm{nc}\Big[(I_\text{x}^{j2}-I_\text{y}^{j2})\cos^2\theta
+(I_\text{x}^jI_\text{z}^j+I_\text{z}^jI_\text{x}^j)\sin2\theta\Big],
\end{split}
\end{align*}
where $A_\mathrm{nc}=A_\mathrm{c}B_\text{Q}/\omega_\text{n}$, describing an effective non-collinear electron--nucleus interaction. This correction term, however, also describes a first-order energetically forbidden nuclear transition. Dressing the Hamiltonian with $V_\text{Q}'$ replaces it by a correction to the electronic driving term,
\begin{align*}
\begin{split}
  V_\text{Q}''= -2\Omega S_\text{y}\sum_j&\frac{A_\mathrm{nc}}{\omega_\text{n}}\Big[\frac{1}{2}(I_\text{x}^jI_\text{y}^j+I_\text{y}^jI_\text{x}^j)\cos^2\theta \\
&+(I_\text{z}^jI_\text{y}^j+I_\text{y}^jI_\text{z}^j)\sin2\theta\Big],
\end{split}
\end{align*}
which describes higher-order simultaneous nuclear--electron transitions induced by the Raman drive.

To evaluate the strength of sideband transitions, we evaluate matrix elements of $V_\text{Q}''$ with respect to the collective nuclear $I_\text{z}$-eigenstates, $\ket{I_\text{z}}$, we find the matrix elements
\begin{align}
\label{eq:nuc-matrix-element}
\begin{split}
  \mel{I_\text{z}'}{V_\text{Q}''}{I_\text{z}} =-2S_\text{y} \frac{\Omega A_\mathrm{nc}}{\omega_\text{n}}&\mathcal{D}_{I_\text{z},I_\text{z}'} \Big\{\sin2\theta\qty[\delta_{I_\text{z}',I_\text{z}+1}+\delta_{I_\text{z}',I_\text{z}-1}] \\  &+\frac{\cos^2\theta}{2}\qty[\delta_{I_\text{z}',I_\text{z}+2}+\delta_{I_\text{z}',I_\text{z}-2}]\Big\},
\end{split}
\end{align}
where $\mathcal{D}_{I_\text{z},I_\text{z}'}$ is an enhancement factor accounting for the degeneracy of the transition. The terms in Eq.~\eqref{eq:nuc-matrix-element} proportional to $\sin 2\theta$ describe the $\Delta I_z = \pm 1$ sideband transitions, and the terms proportional to $\cos^2\theta$ describe the $\Delta I_z = \pm2$ transitions. Assuming that the nuclear bath is unpolarized and thus $I_\text{z},I_\text{z}'\simeq 0$, we find for spin-3/2 nuclei 
\begin{align}
\label{eq:degeneracy}
  \abs{\mathcal{D}_{I_\text{z},I_\text{z}'}}\simeq \sqrt{3N/4}.
\end{align}

\subsection{Theory for sideband spectrum}

To model the excitation spectrum of the electron spin in the presence of the nuclear sideband transitions, we assume that the initial nuclear density operator can be written as a classical mixture of $\ket{I_\text{z}}$-states described by the cooled Overhauser distribution, $p(I_\text{z})$. In each realisation of the ensemble, $\ket{I_\text{z}}$, we expand the nuclear state on the relevant five-dimensional subspace $\{\ket{I_\text{z}},\ket{I_\text{z}\pm 1},\ket{I_\text{z}\pm 2}\}$ and calculate the dynamics under the approximation \eqref{eq:degeneracy} using the master equation for the combined electron-nuclear system, 
\begin{align}
\label{eq:master-eq}
\begin{split}
  \pdv{\rho(t|I_z)}{t}&=i[\rho(t|I_\text{z}),H]+\sum_{I_\text{z}'=I_\text{z}-2}^{I_\text{z}+2} \Gamma_\text{n} L(\dyad{I_\text{z}'})\\&+\frac{1}{T_2}L(S_\text{z}),
  \end{split}
\end{align}
where $\rho(t|I_\text{z})$ is the density operator conditional on the initial nuclear state being $I_\text{z}$, $L(a)=a\rho(t|I_\text{z})a^\dagger -\frac{1}{2}\{a^\dagger a,\rho(t|I_\text{z})\}$ is the Lindblad operator, and $\Gamma_\text{n}$ is broadening of the collective nuclear states due to quadrupolar fine structure structure, $\Gamma_\text{n}=2(1+\alpha)\abs{\Delta_\text{Q}}$. Here, $\alpha$ is the relative variation of $\Delta_\text{Q}$ in the nuclear ensemble due to inhomogeneous strain, which is estimated to be $\sim 80\%$, based on \cite{Bulutay2012b}. The last term in \eqref{eq:master-eq} accounts for electronic dephasing induced by nuclear diffusion. The ensemble density operator, $\chi$, is then calculated by averaging the conditional density operator $\rho(t|I_\text{z})$ over the different initial configurations $\chi(t)=\int\dd{I_\text{z}}p(I_\text{z})\rho(t|I_\text{z})$.

Fig.~3b (main text) shows the theoretically predicted excitation map resulting from the master equation, \eqref{eq:master-eq}. The quadrupolar parameters are fitted to the experiment (Fig.~3e main text), from which we obtain: $B_\text{Q}=1.7\mathrm{\: MHz},\; \theta=20.4^\circ$. These fitted values are consistent with parameters for As nuclei from literature\cite{Bulutay2012b}. Other parameters are taken from the experimental configuration, $\omega_\text{n}/2\pi=7.22\mathrm{\: MHz/T}, B=3$~T, $\Omega/2\pi=3.3\mathrm{\: MHz},\; T_2=1.5\mathrm{\: \mu s}$. 

In relation to the sideband strength relative to the carrier Rabi frequency, $\eta$, these parameters yield the first sideband strength $\eta_1 = (A_\text{nc}/\omega_\text{n})(\sqrt{3N/4})(\sin 2 \theta)=0.1$ and the second sideband strength $\eta_2 = (A_\text{nc}/\omega_\text{n})(\sqrt{3N/4})(\cos^2 \theta/2)=0.14$.

\section{More details on fitting the electron resonance spectrum}

The spectrum in Fig. 3f (main text) is fitted with a sum of five Gaussian functions $\mathcal{G}(A,\delta,\sigma)$ of amplitude $A$ (population), detuning $\delta$ (MHz), and standard deviation $\sigma$ (MHz), appearing as a dashed line that agrees closely with the data. The result of the fit is: $\mathcal{G}(0.40(1),0.0(2),7.5(3))+\mathcal{G}(0.27(1),22.0(3),7.2(4))+\mathcal{G}(0.27(1),-22.6(3),7.1(4))+\mathcal{G}(0.10(1),46.3(9),9.6(9))+\mathcal{G}(0.13(1),-47.8(9),11.9(9))$.

\section{Raman cooling}
\subsection{Cooling model}
\subsubsection{Two-level limit}

We consider two fields with orthogonal polarization driving a three-level system: a $V$-polarized field driving the $V$-polarized exciton transition $|\uparrow\rangle$ to the trion state $|\Uparrow \uparrow \downarrow\rangle$ with resonant Rabi frequency $\Omega_\text{V}$ and detuning $\Delta_\text{V}$, and an $H$-polarized field driving the $H$-polarized exciton transition $|\downarrow\rangle$ to the same trion state $|\Uparrow \uparrow \downarrow\rangle$ with resonant Rabi frequency $\Omega_\text{H}$ and detuning $\Delta_\text{H}$. In the limit where these fields do not populate the trion excited state, $\Omega_\text{H,V}^2/\Delta_\text{H,V}^2 \ll 1$, we can ignore the excited state contribution and reduce the driven three-level system to a driven two-level system split by the electron Zeeman energy $\omega_\text{e}$, where the two-photon transition between the electronic states $|\uparrow\rangle$ and $\downarrow\rangle$ has a resonant Rabi frequency $\Omega = \Omega_\text{H} \Omega_\text{V}/2\Delta$, $\Delta = (\Delta_\text{H} + \Delta_\text{V})/2$, and a detuning from electron spin resonance of $\delta = \Delta_\text{V} - \Delta_\text{H} - \omega_\text{e}$.

In the presence of an additional field with resonant Rabi frequency $\Omega_\text{p}$ performing resonant optical pumping on the $H$-polarized transition, the $|\downarrow\rangle$ state acquires an effective linewidth that is the inverse of its lifetime under optical pumping:
\begin{equation*}
\begin{split}
%\Gamma = \frac{\Gamma_0}{4} \frac{2\left(\frac{\Omega_\text{p}}{\Gamma_0}\right)^2}{1+\frac{\Omega_\text{p}}{\Gamma_0}\right)^2}
\Gamma = \frac{\Gamma_0}{4} \frac{2\left(\Omega_\text{p}/\Gamma_0\right)^2}{1+2\left(\Omega_\text{p}/\Gamma_0\right)^2} \text{,}
\end{split}
\end{equation*}
\noindent
where $\Gamma_0 \approx 150$~MHz, is the natural linewidth of the trion excited state. The optical pumping field Rabi frequency $\Omega_\text{p}$ can thus be used to tune the excited state linewidth of the two-level system in the range $[0,\Gamma_0/4]$.

We can hereon treat scattering in the two-level system as per the textbook formula, given a driving field with Rabi frequency $\Omega$, excited state with linewidth $\Gamma$, and dephasing rate $\Gamma_2$. In relation to the main text, this gives then the stimulated Raman scattering rate on the central electron spin transition at a detuning $\delta$:
\begin{equation*}
\begin{split}
W(\delta) = \frac{\Gamma}{2} \frac{(\Omega^2/\Gamma\Gamma_2)}{1+(\Omega^2/\Gamma\Gamma_2) + \left(\delta/\Gamma_2\right)^2} \text{,}
\end{split}
\end{equation*}
\subsubsection{Cooling}

The sideband processes are best represented in a ladder of states, where we consider the electron spin states dressed by the nuclear states of polarization $I_\text{z} = [-NI,NI]$ (for spin-$I$ nuclei), as in Fig.~1 (main text). In this simple picture, $V_\text{Q}''$ allows transitions which change nuclear spin polarization by one (or two) nuclear spin flip, in either direction. Towards cooling, which results in a polarization which deviates from its steady-state over an energy scale well below the nuclear Zeeman energy, we consider that the $I_z \pm 1$ sideband largely dominates the process, and its coupling strength is a fraction $\eta =  (\sqrt{NI/2})( A_\text{nc}/N\omega_\text{n})(\sin 2\theta)$ (for spin-$I$) of the coupling on the central transition.

Combining these sideband processes within our Raman-driven two-level system where one electron spin state relaxes to the other, we obtain two types of rates that change the nuclear spin polarization, drift rates $W_\pm$ arising from stimulated sideband transitions, and a diffusion rate $\Gamma_\text{nc}$ arising from spontaneous sideband transitions, each occuring at a maximum rate of $\eta^2 \Gamma/2$ with a spectral response function defined by the polarization-dependent Raman scattering rate $W(\delta,I_z)$:
\begin{eqnarray*}
\begin{split}
W_\pm(\delta,I_z) &= \eta^2 \frac{\Gamma}{2} \frac{(\Omega^2/\Gamma\Gamma_2)}{1+(\Omega^2/\Gamma\Gamma_2) + \left(\Delta_\pm(\delta,I_z)/\Gamma_2\right)^2}\\
\Gamma_\text{nc}(\delta,I_z) &= \eta^2 \frac{\Gamma}{2} \frac{(\Omega^2/\Gamma\Gamma_2)}{1+(\Omega^2/\Gamma\Gamma_2) + \left((\delta-A_\text{c} I_\text{z})/\Gamma_2\right)^2}\\
\end{split}
\end{eqnarray*}
\noindent
Where the effective detuning $\Delta_\pm(\delta) = (\delta-A_\text{c}(I_\text{z}\pm1)\mp \omega_\text{n})$, and we take the dephasing rate $\Gamma_2 = (\Gamma/2 + 1/T_2)\sqrt{1+ 2\left(\Omega_\text{p}/\Gamma_0\right)^2} + \Delta \omega_\text{n}$ to be determined by the electronic excited state linewidth $\Gamma$, the homogeneous dephasing time of the electron spin resonance $T_2$, power broadened by the optical pumping saturation $2\left(\Omega_\text{p}/\Gamma_0\right)^2$, and by the inhomogeneous broadening $\Delta \omega_\text{n}$ of the nuclear Zeeman energies arising primarily from the multiple nuclear species partaking in the process. 

For a fixed two-photon detuning $\delta = 0$, the evolution of the nuclear polarization is given by:
\begin{equation*}
\begin{split}
\frac{dI_\text{z}}{dt} = W_+(I_\text{z})(1-\frac{I_z}{NI}) - W_-(I_\text{z})(1+\frac{I_z}{NI})- \Gamma_d(I_\text{z})\frac{I_z}{NI}
\end{split}
\label{Eq:fb}
\end{equation*}
Where $\Gamma_\text{d}(I_z) = \Gamma_\text{nc}(I_z) + \Gamma_\text{em}$ is the total diffusion term composed of a polarization-dependent optical diffusion $\Gamma_\text{nc}$ as defined above, and a constant rate non-optical electron mediated diffusion term $\Gamma_\text{em}$ which we report on in a later section of this supplementary.

We can re-arrange this rate equation:
\begin{equation*}
\begin{split}
\frac{dI_\text{z}}{dt} &= -\left(W_+(I_\text{z}) + W_-(I_\text{z}) + \Gamma_d(I_\text{z})\right)\\
&\left(\frac{I_z}{NI}-\frac{W_+(I_\text{z}) - W_-(I_\text{z})}{W_+(I_\text{z}) + W_-(I_\text{z}) + \Gamma_d(I_\text{z})}\right)\\
\frac{dI_\text{z}}{dt} &= -\frac{\Gamma_\text{tot}}{NI}\left(I_z-NIs^{(1/2)}(I_z)\right)\text{,}
\end{split}
\label{Eq:fb}
\end{equation*}
\noindent
where, following Yang \& Sham\cite{Yang2013b}, we have defined $\Gamma_\text{tot} = W_+(I_\text{z}) + W_-(I_\text{z}) + \Gamma_d(I_\text{z})$, and have made the association that $s^{(1/2)}(I_z) = (W_+(I_\text{z}) - W_-(I_\text{z}))/\Gamma_\text{tot}$ is the steady state fractional polarization for spin-$1/2$ nuclei. This acts then as proportional to the cooling function $f(I_\text{z}) = NI s^{(1/2)}(I_\text{z})$ as per the formula in the main text. We make the approximation that the nuclear polarization is small, $|I_\text{z}| \ll N$, which is where we conduct our experiments, and as such we can approximate the steady-state nuclear polarization for an arbitrary spin-$I$ as:
\begin{equation*}
\begin{split}
s(I_\text{z}) \approx \frac{2}{3}(I+1)\frac{W_+(I_\text{z}) - W_-(I_\text{z})}{W_+(I_\text{z}) + W_-(I_\text{z}) + \Gamma_d(I_\text{z})}
\end{split}
\label{Eq:fb}
\end{equation*}
Steady-state is then simply obtained by the numerically solvable self-consistent equation:
\begin{equation*}
\begin{split}
s(I_0) =  I_0/NI
\end{split}
\label{Eq:fb}
\end{equation*}
We note here that the gradient $f'(I_0)$, which we relate to the damping coefficient, depends intimately on the relationship between $\omega_\text{n}$, $A_\text{c}$, and $\Gamma$ through the spectral function $\Delta_{\pm}/\Gamma_2$. Two competing demands exists on $\Gamma$. First, the simple condition that some sideband scattering is required near the stable point $I_0$ in order for good Raman cooling to occur means that the Raman rate $\Omega$ and linewidth $\Gamma$ must be comparable to the nuclear Zeeman energy $\omega_\text{n}$; this means making $\Gamma/\omega_\text{n} \approx 1$. Secondly, the spectral selectivity over the Overhauser shift $2 A_\text{c} I_\text{z}$, which improves narrowing around a given value of $I_z$, depends on the ratio $A_\text{c}/\Gamma$, which needs to be large for good cooling. These two competing effects lead to an optimal value of $\Gamma$, which is ultimately determined by just how much scattering is needed near the stable point $I_0$ to beat non-optical diffusion $\Gamma_\text{em}$.

Linearising $f(I_\text{z})$ around the stable point $I_0$, it is possible to obtain the steady-state distribution $p(I_\text{z})$ under a Fokker-Planck treatment of the diffusion equation, and to obtain an analytical expression for the steady-state reduction in the width of the thermal distribution $\Delta I_\text{z,th}^2$:
\begin{equation*}
\begin{split}
\frac{\Delta I_\text{z}^2}{\Delta I_\text{z,th}^2} = \frac{1-\left(\frac{3}{2(I+1)}\frac{I_0}{NI}\right)^2}{1-\frac{2(I+1)}{3}f'(I_0)}
\end{split}
\label{Eq:fb}
\end{equation*}
Hence, it is clear that for any reduction of fluctuations around $I_0$, we require $|f'(I_0)|>0$. The above analytical result forms the basis for our theoretical cooling predictions in Fig. 2 (main text). In our measurements we obtain the ratio $\Delta I_\text{z}^2/\Delta I_\text{z,th}^2$ by exploiting the straightforward relationship between $\Delta I_\text{z}^2$ and $T_2^*$ (obtained from Ramsey measurement of the electron spin splitting):
\begin{equation*}
\begin{split}
\frac{\Delta I_\text{z}^2}{\Delta I_\text{z,th}^2} = \left(\frac{T_{2,\text{th}}^*}{T_{2}^*}\right)^2
\end{split}
\end{equation*}
%

% FIGURE S3
\begin{figure}[!b]
  \centering
  \includegraphics[width=\columnwidth]{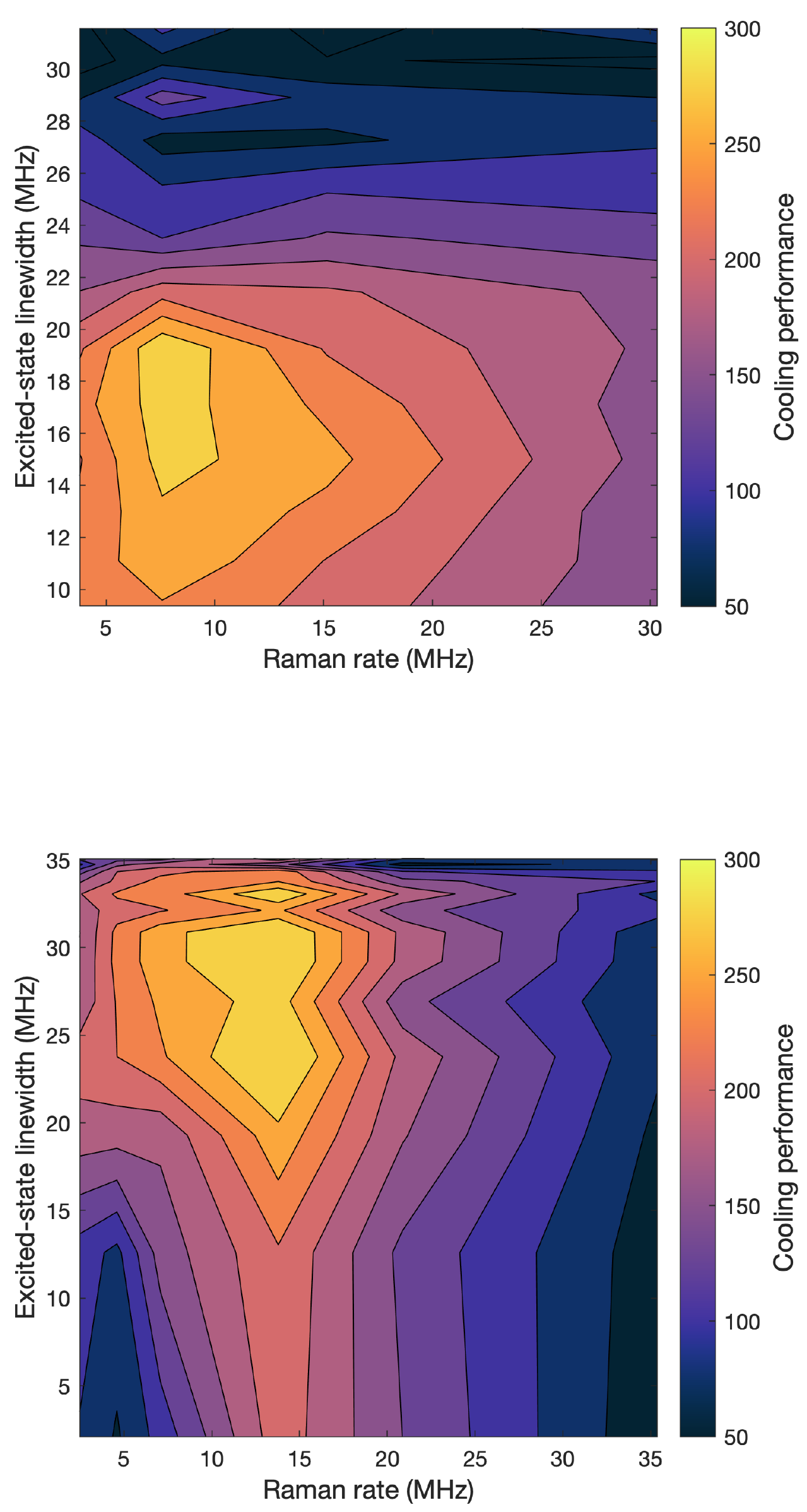}
  \caption{\textbf{Raman cooling at 3T and 6T} \textbf{(a)} Experimental map of the cooling performance as a function of Raman rate and excited state linewidth at a magnetic field of 3T \textbf{(b)} The same at a magnetic field of 6T.}
\end{figure}
\subsection{Further measurements}
\subsubsection{Cooling measurements at $3$~T and $6$~T}

In Fig.~2a (main text), we reported our cooling performance measurements as a function of both Raman rate and excited state linewidth performed at a magnetic field of $5$~T. The same measurements were performed at $3$~T and $6$~T, which we show here in Fig.~S3. The optimal cooling performance and the corresponding Raman rate and linewidth are those reported in Fig.~2d,e (main text).

\subsubsection{Note on measurement at $4$~T}

At $4$~T, we only varied the excited state linewidth $\Gamma$ near the interpolated optimal value of $\Omega$, which gives a good approximation of the optimal excited state linewidth within our measurement error. As a result we present in Fig.~2d (main text) an optimal excited state value but not an optimal Raman rate at this magnetic field.

\subsubsection{Nuclear spin diffusion: relaxation of the probability distribution variance}

In the absence of cooling ($W_\pm = 0$), the nuclear spin distribution relaxes back to its equilibrium distribution giving us information on the non-optical nuclear spin diffusion rate $\Gamma_\text{em}$ that plays an important role in limiting the cooling efficiency at all fields. In Fig.~S4, we show a measurement performed at $3$~T of the nuclear polarization fluctuations $\Delta I_\text{z}^2$, normalized by their thermal value $5N/4$, as they relax from their cold nonequilibrium steady-state to their thermal state in the absence of cooling, as a function of delay time between cooling and Ramsey measurement of the fluctuations\cite{Ethier-Majcher2017b}.

The relaxation is fitted well by an exponential model for which the asymptotic behaviour is fixed $4\Delta I_\text{z}^2/5N \rightarrow 1$, as we know from measuring the electron coherence at thermal equilibrium\cite{Ethier-Majcher2017b}. The typical relaxation time of the fluctuations $\tau = 41.7\pm1.2$ms is used as the model parameter $1/\Gamma_\text{em}$ which is part of the nuclear depolarization rate $\Gamma_d$ that is used in our theoretical calculations presented in Fig.~2 (main text). This mechanism is only present when an electron is present in the QD, which points towards an electron-mediated mechanism\cite{Latta2011b,Wust2016b,Ethier-Majcher2017b}. In this respect, the diffusion rate ought to be inversely proportional to the electron spin splitting squared $\Gamma_\text{em}(B) \propto B^{-2}$, where $\Gamma_\text{em}(3) = 1/\tau$, and this is the scaling we use to obtain our theory curves in Fig.~2 (main text).

% FIGURE S4
\begin{figure}[!b]
  \centering
  \includegraphics[width=\columnwidth]{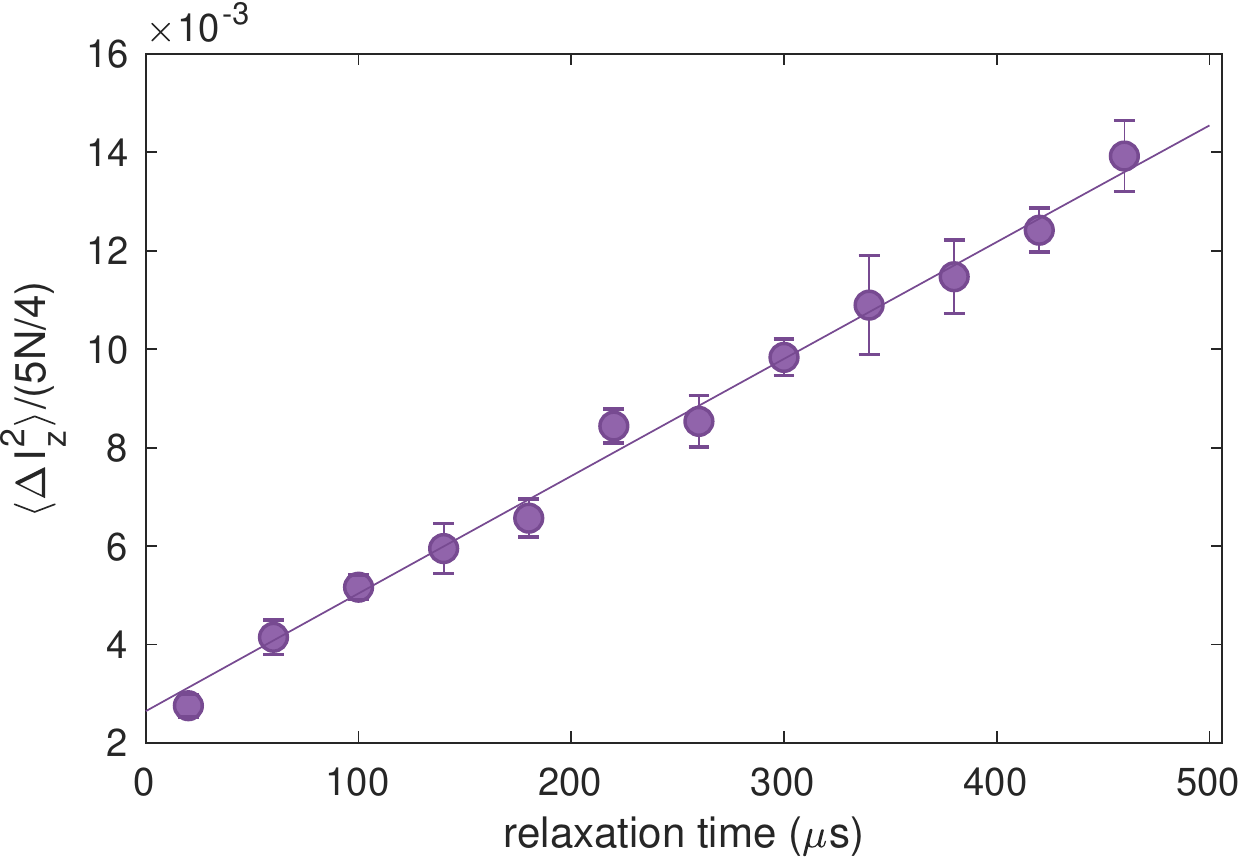}
  \caption{\textbf{Relaxation of the nuclear spin fluctuations} Measurement at 3T of the nuclear polarization variance $\Delta I_\text{z}^2$, normalized by its thermal value $5N/4$, as it relaxes from its cold state to its thermal state in the absence of cooling, as a function of delay (relaxation) time between cooling and Ramsey measurement of the fluctuations. The solid curve is a fit to the model $1-a\exp(-t/\tau)$, for which we obtain $a=0.9974\pm0.0003$ and $\tau = 41.7\pm1.2$~ms.}
\end{figure}

\subsection{Additional notes on global limit to cooling performance}

Even in the absence of non-optical nuclear spin diffusion mechanisms ($\Gamma_\text{em}= 0$, $\Gamma_\text{d} = \Gamma_\text{nc}$), the simple condition that some sideband scattering is required close to the steady state polarization $I_0$ in order to maintain Raman cooling, i.e. $|f'(I_0)|>0$, means that the Raman rate $\Omega$ and linewidth $\Gamma$ must be of the order of the nuclear Zeeman energy $\omega_\text{n}$. The dependence of sideband absorption rate on the polarization $I_\text{z}$ is a function of the Overhauser shift $2A_\text{c}I_\text{z}$, and as a result the damping coefficient $f'(I_0)$ depends on the ratio of the Overhauser shift to the Zeeman energy $A_\text{c}/\omega_\text{n}$. This condition thus entails a limit on the spectral selectivity that is intrinsic to the sideband cooling mechanism we employ: the larger the energy splittings at play, the less efficient the cooling. As a result, the cooling efficiency is inversely proportional to the magnetic field. The resulting field-dependence is shown as a dotted theory line in Fig.~2e (main text) bounding the red shaded region. 

%The cooling efficiency of the Raman process is ultimately limited by spectral broadening and by nuclear spin diffusion, captured by $\Gamma_d$ in eq.~7. 

The field-dependence of the cooling efficiency is accentuated by an explicit dependence of the sideband scattering on magnetic field, i.e. $W_\pm \propto \eta^2 \propto  \omega_\text{n}^{-4}$.  The electron-mediated nuclear spin diffusion mechanism\cite{Latta2011b,Wust2016b} plays a significant role in our system as described in the previous section. This diffusion mechanism decreases the cooling performance at all fields. It has a field dependent rate $\propto B^{-2}$ and, when added to the sideband model, leads to a sharper decrease of cooling efficiency with increasing field, as seen from the high-field behavior of the solid theory line in Fig.~2e (main text).

Lastly, at magnetic fields below 2T, where the nuclear Zeeman eigenstates are strongly mixed by strain and the resulting fast precession of a wide spectrum of nuclear-spin frequencies dephase the electron spin over a time $T_2$\cite{Bechtold2015b,Stockill2016b}, the Raman resonance is homogeneously broadened to a linewidth $\Delta \nu_\text{Q} \sim 1/T_2>20$~MHz and thereby imposes that limit on the width of the nuclear spin distribution that can be prepared with Raman cooling. This leads to the dashed theory line in Fig.~2e (main text) bounding the blue shaded area. 

We summarize our measurements under these cooling limits in Fig.~2e (main text), where it is clear that we reach our system's global temperature optimal at a magnetic field around $3.3$T.

\section{A relation to canonical temperature}

We define the effective temperature of the nuclear spin ensemble as the temperature of a thermal ensemble that would feature the same polarization fluctuations $\Delta I_z^2$ as those we measure. We make a number of assumptions to relate the measured system properties to a temperature:
\begin{itemize}
\item the energy of the system is characterized by $I_\text{z}$ (i.e. we neglect the quadrupole shifts of the collective state)
\item we neglect the presence of multiple nuclear isotopes and take the nuclear Zeeman energy of As (spin-$3/2$), which has the lowest Zeeman splitting, and therefore provides a lower bound on the temperature
\item we consider that the average polarization $I_\text{z}$ probed by the electron spin is representative of the nuclear polarization of the QD, which is only exactly true for homogeneous coupling
\end{itemize}

We define the partition function of the nuclear spin system expressed in the collective $I_z$ basis, where we account for the degeneracy $g(I_z)$ of each state:
\begin{equation*}
\begin{split}
Z(\beta) = \sum_{I_z=-3N/2}^{3N/2} g(I_z) \exp(-\beta I_z)\text{,}
\end{split}
\end{equation*}
\noindent
where $\beta = \hbar \omega_\text{n} / k_\text{B} T$. The degeneracy term $g(I_z)$ is a sum of binomial coefficients corresponding to the number of ways a state $I_z$ can be thermally occupied accounting for individual spin-$3/2$ structures that contain four internal states\cite{Wesenberg2002b}. As an example, for $I_z = -3N/2 + 2$ (i.e. two nuclear spin flips away from maximal polarization), this can occur with two separate spins increasing their internal energy by one unit, or a single spin increasing its internal energy by two units. Here we show this degeneracy factor in order of increasing unit jumps: 
\begin{equation*}
\begin{split}
g(-3N/2) &= 1\\
g(-3N/2+1) &= \left( {\begin{array}{cc}
   N \\
   1 \\
  \end{array} } \right)\\
  g(-3N/2+2) &= \left( {\begin{array}{cc}
   N \\
   2 \\
  \end{array} } \right) + \left( {\begin{array}{cc}
   N \\
   1 \\
  \end{array} } \right)\\
  g(-3N/2+3) &= \left( {\begin{array}{cc}
   N \\
   3 \\
  \end{array} } \right) + \left( {\begin{array}{cc}
   N \\
   1 \\
  \end{array} } \right) \left( {\begin{array}{cc}
   N-1 \\
   1 \\
  \end{array} } \right)\\
  & + \left( {\begin{array}{cc}
   N \\
   1 \\
  \end{array} } \right)\\
  &...
\end{split}
\end{equation*}
What we deduce from this sequence is that the term with single unit jumps (i.e. the first term) dominates the count by a factor $O(N)$ for high polarization $I_z \lesssim -N$ (for which this first term reaches a maximum), and thus serves as a good approximation when considering the low temperature regime $\beta > 1$. We thus approximate the partition function as:
\begin{equation*}
\begin{split}
Z(\beta) \approx \sum_{I_z=-3N/2}^{-N}  \left( {\begin{array}{cc}
   N \\
   I_z+3N/2 \\
  \end{array} } \right) \exp(-\beta I_z)\text{,}
\end{split}
\end{equation*}
\noindent
where we have truncated the sum at $I_z = -N$, which yields a sufficient number of terms in the sum for a low temperature approximation. Note that this result is exact for spin-$1/2$ particles.

From the partition function $Z(\beta)$, it is then a simple matter to calculate the moments for a thermal distribution:
\begin{equation*}
\begin{split}
\langle I_z^k \rangle = \frac{(-1)^k}{Z(\beta)}\frac{\partial^k Z(\beta)}{\partial \beta^k}
\end{split}
\end{equation*}
\noindent
and most importantly the variance $\Delta I_z^2$:
\begin{equation*}
\begin{split}
\Delta I_z^2 &= \langle I_z^2 \rangle - \langle I_z \rangle^2\\
&= \frac{-1}{Z(\beta)}\frac{\partial^2 Z(\beta)}{\partial \beta^2} - \left(\frac{1}{Z(\beta)}\frac{\partial Z(\beta)}{\partial \beta}\right)^2
\end{split}
\end{equation*}
We can plot the mean polarization $|\langle I_z \rangle |$ and the fluctuations $\Delta I_z^2$ as a function of $\beta$, shown in Fig.~S5. From this plot, it becomes clear that any significant  reduction of fluctuations $\Delta I_z^2 \ll N$ is accompanied by a reduction of temperature below the system's defining energy scale, i.e. the nuclear Zeeman splitting $\omega_\text{n}$; this corresponds to $\beta > 1$. Our maximum measured cooling performance $(5N/4)/\Delta I_z^2 = 400$ can also be seen as the reduction of fluctuations to a level of $\Delta I_z^2 = (5N/4)/400 \approx 100$. From our theory curves in Fig.~S5, this corresponds to a thermal ensemble with $\beta = 5.5$ and a polarization around $99.9\%$. There, the equivalent temperature is $T = (\hbar \omega_\text{n}/k_\text{B}) / 5.5 \approx 200 \mu$K.

% FIGURE S5
\begin{figure}[!t]
  \centering
  \includegraphics[width=\columnwidth]{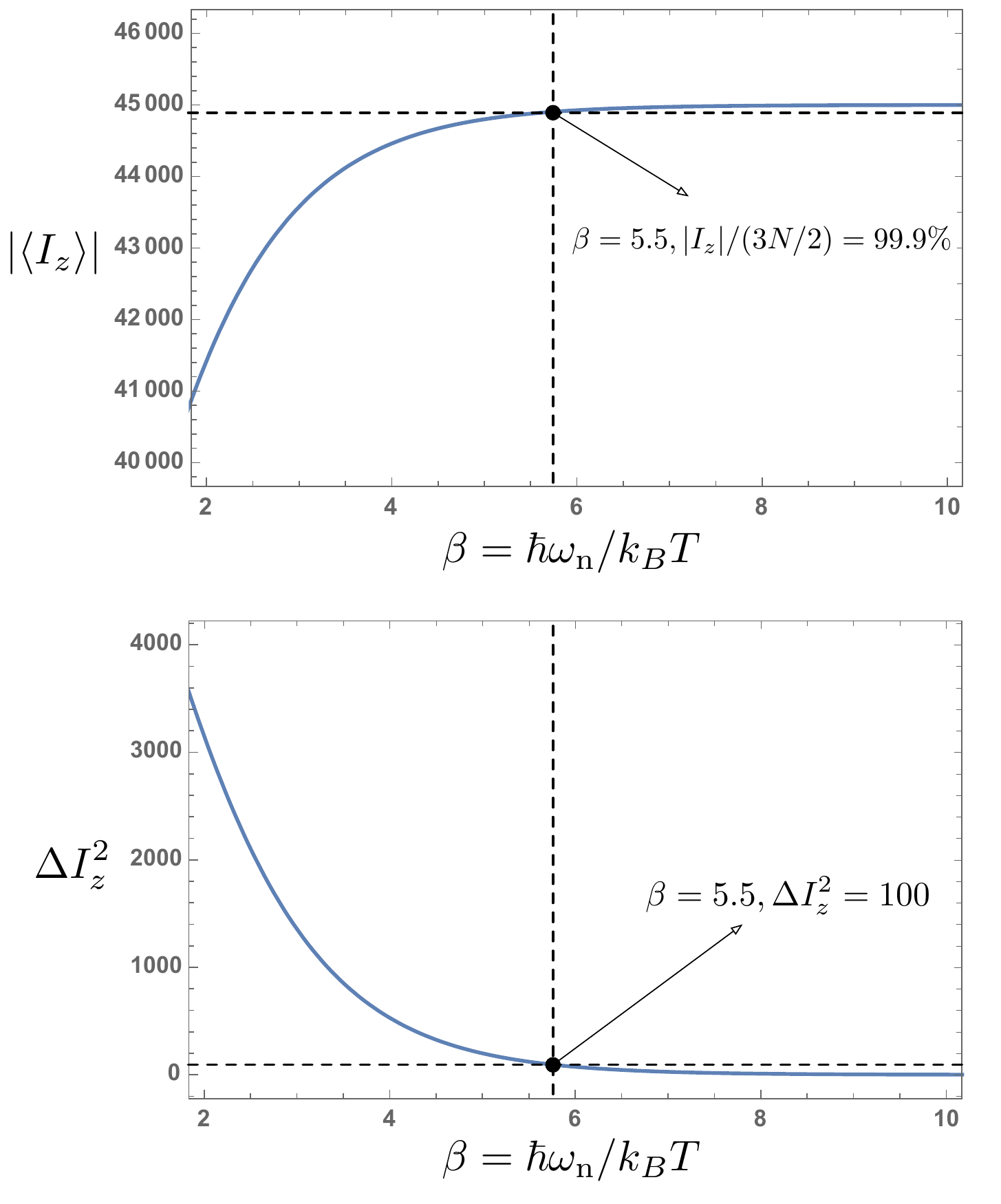}
  \caption{\textbf{Mean polarization and fluctuations as a function of inverse temperature.} Curves are calculated from a canonical ensemble of non-interacting spin-$3/2$ nuclei, in the collective polarization $I_z$ basis, as a function of inverse temperature $\beta = \hbar \omega_\text{n} /k_\text{B} T$. Dashed lines mark values corresponding to our maximum measured cooling performance, $\Delta I_z^2 = (5N/4)/400 \approx 100$.}
\end{figure}

\section{Nuclear magnon vs electron population}

In Fig.~4 (main text), our theory curves showed what we measure in our experiment: the electron excited state population summed over all nuclear states. In our simulations, we can however readout the particular nuclear state $|I_\text{z} - 2\rangle$ that we target with our sideband drive. Here we show as Fig.~S6 a version of Fig.~4 (main text), where the population of the target nuclear state is shown alongside the main theory curve and the data.

% FIGURE S6
\begin{figure}[!b]
  \centering
  \includegraphics[width=\columnwidth]{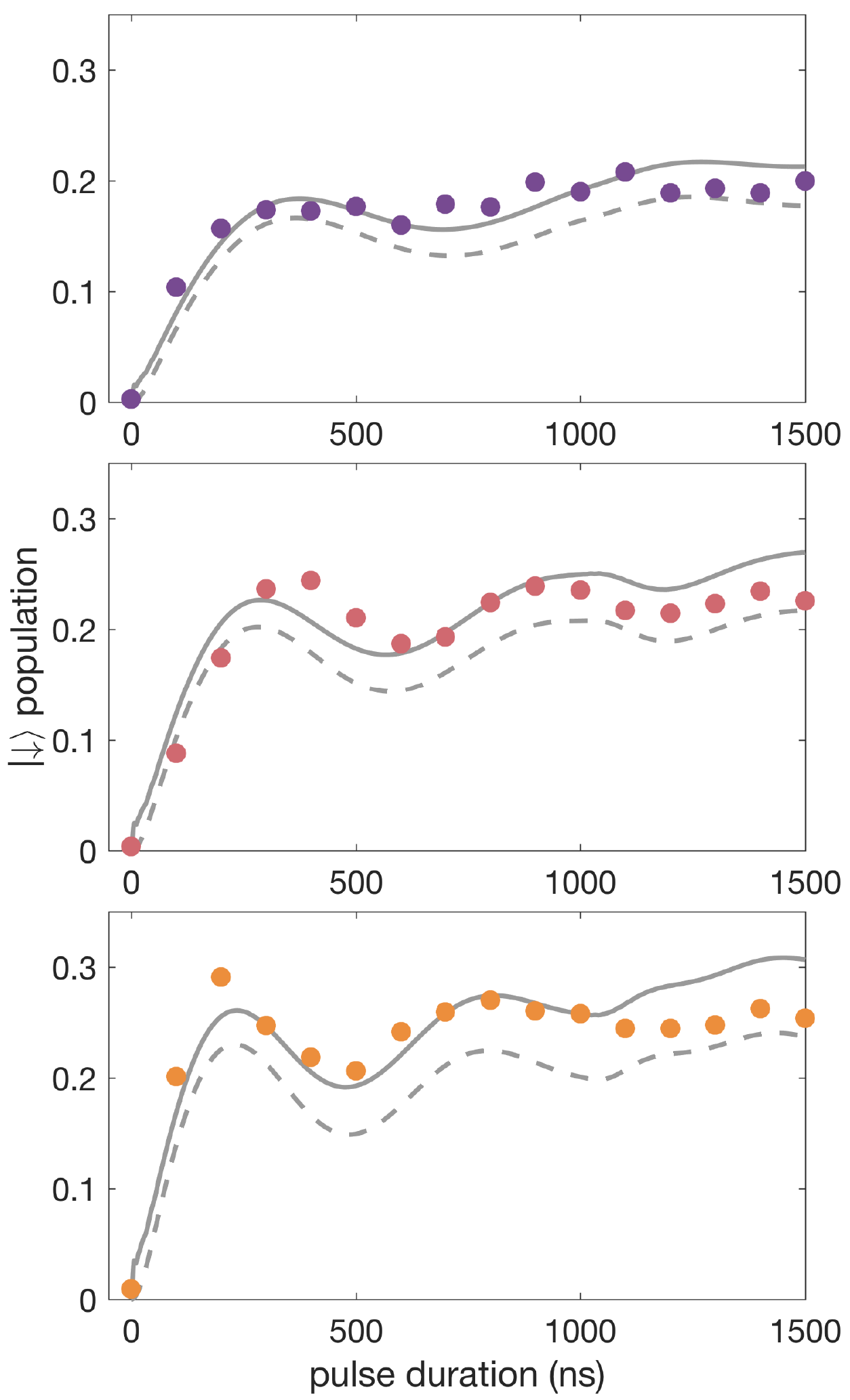}
  \caption{\textbf{Coherent electron-magnon exchange} Electronic excited state $|\downarrow\rangle$ population, measured after a Rabi pulse of $\tau$ at a detuning $\delta =-2\omega_\text{n} = -52$~MHz, for a magnetic field of $3.5$~T, measured for carrier Rabi frequencies $\Omega = 7,9,12$~MHz (top to bottom). Solid curves are the corresponding theoretical calculations of identical carrier Rabi frequencies. The dashed curves are the simulated population transferred only to state $|\downarrow,I_\text{z}-2\rangle$. }
\end{figure}

\section{Summary of model values}

\begin{table}[!!t]
\small
\begin{center}
\caption{Summary of directly measured experimental parameters (single parameter fits) }
\begin{tabular}[b]{ | c || c | }
\hline
Wavelength & $950$~nm\\ \hline
Electron $g_\text{e}\mu_\text{B}$ & $6.3$~GHz/T \\ \hline
Inhomogeneous dephasing time $T_{2,th}^*$ & $1.7$~ns\\ \hline
Homogeneous dephasing time $T_2$ ($2$~T) & $20$~ns\\ \hline
Homogeneous dephasing time $T_2$ ($3$~T) & $500$~ns\\ \hline
Homogeneous dephasing time $T_2$ ($5$~T) & $2000$~ns\\ \hline
Sideband coupling $\eta$ at $3.5$~T (Fig.~4)  & $0.15$ \\ \hline
Nuclear relaxation time $T_{1,\text{n}}$ ($3$~T) & $42$~ms \\ \hline

\end{tabular}
\label{table}
\end{center} 
\end{table}

\begin{table}[!hbp]
\small
\begin{center}
\caption{Summary of fitted model parameters (multi-parameter fits)}
\begin{tabular}[b]{ | c || c | }
\hline
Number of nuclei $N$ (Fig.~2) &  $3 \times 10^4$ \\ \hline
Trion linewidth $\Gamma_0$ (Fig.~2) & $150$~MHz \\ \hline
$A_\text{c}$ (Fig.~2) & $600$~kHz \\ \hline
Nuclear Zeeman spread $\Delta\omega_\text{n}$ at $3$~T (Fig.~2) &  $10$~MHz \\ \hline
$\eta$ at $3$~T (Fig.~2) & 0.063 \\ \hline
$\eta_1$ at $3$~T (Fig.~3 1st sideband) & 0.10 \\ \hline
$\eta_2$ at $3$~T (Fig.~3 2nd sideband) & 0.14 \\ \hline
Nuclear broadening $\Gamma_\text{n}$ (Fig.~3) & 3.9~MHz \\ \hline
Nuclear broadening $\Gamma_\text{n}$ (Fig.~4) & 0.7~MHz \\ \hline
Homogeneous dephasing time $T_2$ at $3.5$~T(Fig.~4) & $5000$~ns\\ \hline

%average nuclear spin $I$ & 2 \\ \hline
\end{tabular}
\label{table}
\end{center} 
\end{table}

Our models contain a number of parameters whose range of values is well-known from previous studies\cite{Bulutay2012b,Hogele2012b,Urbaszek2013b,Stockill2016b}, but whose exact value is determined by fitting our data in several independent experiments.

The cooling experiments of Fig.~2  (main text) are fitted with our single-species Raman cooling model described earlier in this SI. We fix the nuclear Zeeman splitting to be that of As, corresponding to a gyromagnetic ratio of $7.22$~MHz/T, and the spin to $I=3/2$. We introduce a nuclear Zeeman broadening free parameter $\Delta \omega_\text{n}$ representing the effective spread of Zeeman energies arising from the contribution of other species and from quadrupolar interactions, which acts as an inhomogeneous broadening term in the Raman scattering rate. The other free parameters of this model are the number of nuclei $N$, the collinear hyperfine constant $A_\text{c}$, the noncollinear hyperfine constant $A_\text{nc}$, and the trion excited state linewidth $\Gamma_0$. While these parameters are ``free", we emphasize that they are tied to independently measured values, from both our own measurements and from literature, and are thus formally tied to specific ranges. $\Delta \omega_\text{n}$ should be no less than the Zeeman energy broadening of $\sim 3$~MHz arising from strain inhomogeneities for a single species, and no more than these inhomogeneities added to the spread of species Zeeman energies, $\sim 30$~MHz (at $3$~T). The number of nuclei $N$ should be in the range $10^4 - 10^6$\cite{Urbaszek2013b}. The total Overhauser shift $N I A_\text{c}$ should be no less than for GaAs $16.3$~GHz, and no more than for InAs $38.1$~GHz\cite{Urbaszek2013b}. The noncollinear hyperfine constant $A_\text{nc}/A_\text{c}$ should be in the range $0.01-0.08$ based on previous measurements and calculations\cite{Bulutay2012b,Hogele2012b}. The trion state lifetime, known to be $\sim$$1$~ns\cite{Urbaszek2013b}, corresponds to linewidths $100-400$~MHz.

The collinear hyperfine interaction product $A_\text{c}\sqrt{N_\text{t}}$ is tied to the inhomogeneous dephasing time $T_2^*$ of the electron\cite{Bechtold2015b,Stockill2016b} in the absence of any cooling; $T_2^* = \left[(2/3)N_\text{t}\langle A_\text{c}^2 I( I+1)\rangle\right]^{-1/2}$. $\langle \rangle$ is the species-averaged value of individual nuclear spin $I$ and hyperfine constant $A_\text{c}$, where we take the Indium concentration to be $50\%$ as per previous measurements on quantum dots from the same wafer\cite{Stockill2016b}. Note that while not all species may contribute to the cooling equally, all species will contribute to the thermal fluctuations of the ensemble. As such, all $N_\text{t}$ spins in the QD should be counted here. We measure $T_2^*=1.7$~ns, which gives the product $A_\text{c}\sqrt{N_\text{t}} = 190$~MHz. The free parameter value of $A_\text{c}\sqrt{N}$ we arrive at from fitting our cooling data (Fig.~2 main text) with our model is $104$~MHz. Within our model, $N$ is the fitted number of spins that partake in the cooling process, which we know to be effectively smaller than the total number of spins of all species $N_\text{t}$. This discrepancy could be thus explained if indeed $N/N_\text{t} = 33\%$ of spins took part in the cooling.

The fitted values of the total Overhauser shift $NIA_\text{c} = 27$~GHz agrees well with an InGaAs QD with $50\%$ In. The fitted non-collinear hyperfine constant $A_\text{nc}/A_\text{c}=0.015$, and the number of spins $3\times 10^4$ match those of previous studies on non-collinear feedback effects in InGaAs QDs\cite{Hogele2012b}.

The trion excited state linewidth, fitted to $\Gamma_0=150$~MHz, sets the absolute scale on the optical parameters $\Omega$ and $\Gamma$, which are measured through optical saturation. It was left a free parameter in order to match the optimal rates measured in Fig.~2 (main text).

The noncollinear hyperfine interaction constant $A_\text{nc} = A_\text{c} B_\text{Q} /N\omega_\text{n}$ and quadrupolar interaction angle $\theta$ are determined by fitting the sideband spectrum in Fig.~3d (main text), where we measure the ratio of population $\eta_1 = (\sqrt{3N/4}) (A_\text{nc}/\omega_\text{n}) \sin(2\theta)$ on the first sideband ( $I_\text{z} \rightarrow I_\text{z} \pm 1$) and $\eta_2 = (\sqrt{3N/4}) (A_\text{nc}/\omega_\text{n}) \cos^2(\theta)/2$ on the second sideband ( $I_\text{z} \rightarrow I_\text{z} \pm 2$) relative to the principal transition. As noted earlier, this is done with the simplification that the ensemble is made up of a single species, $As$, with a Zeeman energy at $3$~T of $21.6$~MHz. From this fit we obtain $\eta_1 = 0.10$ and $\eta_2=0.14$, or equivalently $B_\text{Q} = 1.7$~MHz, and $\theta = 21^\circ$.

We also measure the value $\eta = (\sqrt{3N/4}) (A_\text{nc}/\omega_\text{n}) \sin(2\theta)$ directly as determined by measuring the ratio of oscillation frequency of electronic population on the second sideband $I_\text{z} \rightarrow I_\text{z} + 2$ relative to the frequency on the principal transition $\Omega_s/\Omega$ (Fig.~4 main text), from which we obtain $\eta = 0.15$.

The discrepency between the sideband-resolved spectra, where $\eta = 0.10-0.15$, and the cooling model, where $\eta = 0.06$, may be attributed to deviations from simple spectral broadening of the sidebands in the case of cooling, whenever multiple species with different coupling strength are corralled to participate in the polarization-changing processes, contrary to our simple single-species model.


\begin{thebibliography}{10}

\bibitem{Amico2008}
L.~Amico, R.~Fazio, A.~Osterloh, V.~Vedral, {\it Rev. Mod. Phys.\/} {\bf 80},
  517 (2008).

\bibitem{Taylor2003a}
J.~M. Taylor, C.~M. Marcus, M.~D. Lukin, {\it Phys. Rev. Lett.\/} {\bf 90},
  206803 (2003).

\bibitem{Kurucz2009}
Z.~Kurucz, M.~W. S{\o}rensen, J.~M. Taylor, M.~D. Lukin, M.~Fleischhauer, {\it
  Phys. Rev. Lett.\/} {\bf 103}, 010502 (2009).

\bibitem{Choi2010}
K.~S. Choi, A.~Goban, S.~B. Papp, S.~J. van Enk, H.~J. Kimble, {\it Nature\/}
  {\bf 468}, 412 (2010).

\bibitem{Abragam1961}
A.~Abragam, L.~C. Hebel, {\it Am. J. Phys.\/} {\bf 29}, 860 (1961).

\bibitem{Stanek2014}
D.~Stanek, C.~Raas, G.~S. Uhrig, {\it Phys. Rev. B\/} {\bf 90}, 064301 (2014).

\bibitem{DeSousa2003}
R.~de~Sousa, S.~{Das Sarma}, {\it Phys. Rev. B\/} {\bf 68}, 115322 (2003).

\bibitem{Pla2012}
J.~J. Pla, {\it et~al.\/}, {\it Nature\/} {\bf 489}, 541 (2012).

\bibitem{Childress2006}
L.~Childress, {\it et~al.\/}, {\it Science (80-. ).\/} {\bf 314}, 281 (2006).

\bibitem{Balasubramanian2009}
G.~Balasubramanian, {\it et~al.\/}, {\it Nat. Mater.\/} {\bf 8}, 383 (2009).

\bibitem{Kalb2017a}
N.~Kalb, {\it et~al.\/}, {\it Science (80-. ).\/} {\bf 356}, 928 (2017).

\bibitem{Khaetskii2002}
A.~V. Khaetskii, D.~Loss, L.~Glazman, {\it Phys. Rev. Lett.\/} {\bf 88}, 186802
  (2002).

\bibitem{Merkulov2002}
I.~A. Merkulov, A.~L. Efros, M.~Rosen, {\it Phys. Rev. B\/} {\bf 65}, 205309
  (2002).

\bibitem{Bluhm2011}
H.~Bluhm, {\it et~al.\/}, {\it Nat. Phys.\/} {\bf 7}, 109 (2011).

\bibitem{Urbaszek2013}
B.~Urbaszek, {\it et~al.\/}, {\it Rev. Mod. Phys.\/} {\bf 85}, 79 (2013).

\bibitem{Tran2018}
M.~C. Tran, J.~M. Taylor, {\it arXiv:1801.04006\/}  (2018).

\bibitem{Stepanenko2006}
D.~Stepanenko, G.~Burkard, G.~Giedke, A.~Imamoglu, {\it Phys. Rev. Lett.\/}
  {\bf 96}, 136401 (2006).

\bibitem{Greilich2007a}
A.~Greilich, {\it et~al.\/}, {\it Science (80-. ).\/} {\bf 317}, 1896 (2007).

\bibitem{Reilly2008}
D.~J. Reilly, {\it et~al.\/}, {\it Science (80-. ).\/} {\bf 321}, 817 (2008).

\bibitem{Xu2009}
X.~Xu, {\it et~al.\/}, {\it Nature\/} {\bf 459}, 1105 (2009).

\bibitem{Vink2009}
I.~T. Vink, {\it et~al.\/}, {\it Nat. Phys.\/} {\bf 5}, 764 (2009).

\bibitem{Bluhm2010}
H.~Bluhm, S.~Foletti, D.~Mahalu, V.~Umansky, A.~Yacoby, {\it Phys. Rev.
  Lett.\/} {\bf 105}, 216803 (2010).

\bibitem{Issler2010}
M.~Issler, {\it et~al.\/}, {\it Phys. Rev. Lett.\/} {\bf 105}, 267202 (2010).

\bibitem{Chow2016}
C.~M. Chow, {\it et~al.\/}, {\it Phys. Rev. Lett.\/} {\bf 117}, 1 (2016).

\bibitem{Onur2016}
A.~R. Onur, {\it et~al.\/}, {\it Phys. Rev. B\/} {\bf 93}, 161204 (2016).

\bibitem{Ethier-Majcher2017}
G.~{\'{E}}thier-Majcher, {\it et~al.\/}, {\it Phys. Rev. Lett.\/} {\bf 119},
  130503 (2017).

\bibitem{Heinzen1990}
D.~J. Heinzen, D.~J. Wineland, {\it Phys. Rev. A\/} {\bf 42}, 2977 (1990).

\bibitem{SuppInfo}
{Supplementary Information available online}.

\bibitem{Bechtold2015}
A.~Bechtold, {\it et~al.\/}, {\it Nat. Phys.\/} {\bf 11}, 1005 (2015).

\bibitem{Stockill2016}
R.~Stockill, {\it et~al.\/}, {\it Nat. Commun.\/} {\bf 7}, 12745 (2016).

\bibitem{Wust2016}
G.~W{\"{u}}st, {\it et~al.\/}, {\it Nat. Nanotechnol.\/} {\bf 11}, 885 (2016).

\bibitem{Monroe1995g}
C.~Monroe, {\it et~al.\/}, {\it Phys. Rev. Lett.\/} {\bf 75}, 4011 (1995).

\bibitem{Phillips1998}
W.~D. Phillips, {\it Rev. Mod. Phys.\/} {\bf 70}, 721 (1998).

\bibitem{Huang2010}
C.-W. Huang, X.~Hu, {\it Phys. Rev. B\/} {\bf 81}, 205304 (2010).

\bibitem{Hogele2012}
A.~H{\"{o}}gele, {\it et~al.\/}, {\it Phys. Rev. Lett.\/} {\bf 108}, 197403
  (2012).

\bibitem{Eble2006}
B.~Eble, {\it et~al.\/}, {\it Phys. Rev. B\/} {\bf 74}, 081306 (2006).

\bibitem{Urbaszek2007}
B.~Urbaszek, {\it et~al.\/}, {\it Phys. Rev. B\/} {\bf 76}, 201301 (2007).

\bibitem{Tartakovskii2007}
A.~I. Tartakovskii, {\it et~al.\/}, {\it Phys. Rev. Lett.\/} {\bf 98}, 026806
  (2007).

\bibitem{Maletinsky2007}
P.~Maletinsky, A.~Badolato, A.~Imamoglu, {\it Phys. Rev. Lett.\/} {\bf 99},
  056804 (2007).

\bibitem{Yang2013}
W.~Yang, L.~J. Sham, {\it Phys. Rev. B\/} {\bf 88}, 235304 (2013).

\bibitem{Latta2011}
C.~Latta, A.~Srivastava, A.~Imamoğlu, {\it Phys. Rev. Lett.\/} {\bf 107},
  167401 (2011).

\bibitem{Dicke1954b}
R.~H. Dicke, {\it Phys. Rev.\/} {\bf 93}, 99 (1954).

\bibitem{Johnson1984}
B.~R. Johnson, {\it et~al.\/}, {\it Phys. Rev. Lett.\/} {\bf 52}, 1508 (1984).

\bibitem{Seewald1997}
G.~Seewald, E.~Hagn, E.~Zech, {\it Phys. Rev. Lett.\/} {\bf 78}, 5002 (1997).

\bibitem{Abdurakhimov2015}
L.~V. Abdurakhimov, Y.~M. Bunkov, D.~Konstantinov, {\it Phys. Rev. Lett.\/}
  {\bf 114}, 226402 (2015).

\end{thebibliography}

\begin{thebibliography}{10}

\bibitem{Urbaszek2013b}
B.~Urbaszek, {\it et~al.\/}, {\it Rev. Mod. Phys.\/} {\bf 85}, 79 (2013).

\bibitem{Bechtold2015b}
A.~Bechtold, {\it et~al.\/}, {\it Nat. Phys.\/} {\bf 11}, 1005 (2015).

\bibitem{Stockill2016b}
R.~Stockill, {\it et~al.\/}, {\it Nat. Commun.\/} {\bf 7}, 12745 (2016).

\bibitem{Press2008b}
D.~Press, T.~D. Ladd, B.~Zhang, Y.~Yamamoto, {\it Nature\/} {\bf 456}, 218
  (2008).

\bibitem{Greilich2009b}
A.~Greilich, {\it et~al.\/}, {\it Nat. Phys.\/} {\bf 5}, 262 (2009).

\bibitem{Onur2018b}
A.~R. Onur, C.~H. van~der Wal, {\it Phys. Rev. B\/} {\bf 98}, 165304 (2018).

\bibitem{Press2010b}
D.~Press, {\it et~al.\/}, {\it Nat. Photonics\/} {\bf 4}, 367 (2010).

\bibitem{Bulutay2012b}
C.~Bulutay, {\it Phys. Rev. B\/} {\bf 85}, 115313 (2012).

\bibitem{Hogele2012b}
A.~H{\"{o}}gele, {\it et~al.\/}, {\it Phys. Rev. Lett.\/} {\bf 108}, 197403
  (2012).

\bibitem{Schrieffer1966b}
J.~R. Schrieffer, P.~A. Wolff, {\it Phys. Rev.\/} {\bf 149}, 491 (1966).

\bibitem{Bravyi2011b}
S.~Bravyi, D.~P. DiVincenzo, D.~Loss, {\it Ann. Phys. (N. Y).\/} {\bf 326},
  2793 (2011).

\bibitem{Yang2013b}
W.~Yang, L.~J. Sham, {\it Phys. Rev. B\/} {\bf 88}, 235304 (2013).

\bibitem{Ethier-Majcher2017b}
G.~{\'{E}}thier-Majcher, {\it et~al.\/}, {\it Phys. Rev. Lett.\/} {\bf 119},
  130503 (2017).

\bibitem{Latta2011b}
C.~Latta, A.~Srivastava, A.~Imamoğlu, {\it Phys. Rev. Lett.\/} {\bf 107},
  167401 (2011).

\bibitem{Wust2016b}
G.~W{\"{u}}st, {\it et~al.\/}, {\it Nat. Nanotechnol.\/} {\bf 11}, 885 (2016).

\bibitem{Wesenberg2002b}
J.~Wesenberg, K.~M{\o}lmer, {\it Phys. Rev. A\/} {\bf 65}, 062304 (2002).

\bibitem{Bulutay2012b}
C.~Bulutay, {\it Phys. Rev. B\/} {\bf 85}, 115313 (2012).

\end{thebibliography}
\end{document}